\documentclass[11pt]{article}

\usepackage{hyperref,hep}
\usepackage{amssymb}
\usepackage{graphicx}
\usepackage{color}
\usepackage{subfigure}

\begin{document}

\baselineskip 6mm
\renewcommand{\thefootnote}{\fnsymbol{footnote}}


\newcommand{\nc}{\newcommand}
\newcommand{\rnc}{\renewcommand}



\newcommand{\tcb}{\textcolor{blue}}
\newcommand{\tcr}{\textcolor{red}}
\newcommand{\tcg}{\textcolor{green}}


\def\be{\begin{equation}}
\def\ee{\end{equation}}
\def\ba{\begin{array}}
\def\ea{\end{array}}
\def\bea{\begin{eqnarray}}
\def\eea{\end{eqnarray}}
\def\nn{\nonumber\\}


\def\ct{\cite}
\def\la{\label}
\def\eq#1{(\ref{#1})}


\def\a{\alpha}
\def\b{\beta}
\def\g{\gamma}
\def\G{\Gamma}
\def\d{\delta}
\def\D{\Delta}
\def\e{\epsilon}
\def\et{\eta}
\def\ph{\phi}
\def\Ph{\Phi}
\def\ps{\psi}
\def\Ps{\Psi}
\def\k{\kappa}
\def\l{\lambda}
\def\L{\Lambda}
\def\m{\mu}
\def\n{\nu}
\def\th{\theta}
\def\Th{\Theta}
\def\r{\rho}
\def\s{\sigma}
\def\S{\Sigma}
\def\ta{\tau}
\def\o{\omega}
\def\O{\Omega}
\def\pr{\prime}


\def\half{\frac{1}{2}}

\def\goto{\rightarrow}

\def\na{\nabla}
\def\grad{\nabla}
\def\curl{\nabla\times}
\def\div{\nabla\cdot}
\def\pa{\partial}
\def\fr{\frac}

\def\bra{\left\langle}
\def\ket{\right\rangle}
\def\lb{\left[}
\def\lc{\left\{}
\def\ls{\left(}
\def\lp{\left.}
\def\rp{\right.}
\def\rb{\right]}
\def\rc{\right\}}
\def\rs{\right)}

\def\vac#1{\mid #1 \rangle}


\def\td#1{\tilde{#1}}
\def\check{ \maltese {\bf Check!}}


\def\Tr{{\rm Tr}\,}
\def\det{{\rm det}}


\def\bc#1{\nnindent {\bf $\bullet$ #1} \\ }
\def\ch {$<Check!>$ }
\def\ss {\vspace{1.5cm}}
\def\text#1{{\rm #1}}
\def\Id{\mathds{1}}

\begin{titlepage}


\vspace{25mm}

\begin{center}
{\Large \bf Holographic Entanglement Entropy with Momentum Relaxation  }

\vskip 1. cm
  {Kyung Kiu Kim$^{a}$\footnote{e-mail : kimkyungkiu@sejong.ac.kr} Chanyong Park$^{b,c,d}$\footnote{e-mail : cyong21@gist.ac.kr}, Jung Hun Lee$^{b,c}$\footnote{e-mail : junghun.lee@gist.ac.kr} and Byoungjoon Ahn$^{e}$\footnote{e-mail : bjahn@yonsei.ac.kr}}

\vskip 0.5cm

{\it $^a\,$ Department of Physics, Sejong University, Seoul, 05006, Korea }\\
{\it $^b\,$Department of Physics and Photon Science, Gwangju Institute of Science and Technology,
Gwangju 61005, Korea}\\
{\it $^c\,$ Asia Pacific Center for Theoretical Physics, Pohang, 790-784, Korea } \\
{\it $^d\,$ Department of Physics, Postech, Pohang, 790-784, Korea }\\
{\it $^e\,$ Department of Physics, College of Science, Yonsei University, Seoul 120-749, Korea }\\

\end{center}

\thispagestyle{empty}

\vskip2cm


\centerline{\bf ABSTRACT} \vskip 4mm

\vspace{1cm}

We studied the holographic entanglement entropy for a strip and sharp wedge entangling regions in momentum relaxation systems. In the case of strips, we found analytic and numerical results for the entanglement entropy and {\color{black} examined the effect of the electric field on the entanglement entropy}. We also studied the entanglement entropy of wedges and confirmed that there is a linear change {\color{black} due to the electric field}. {\color{black} In this case, we showed that the entanglement entropy change is interestingly proportional to the thermoelectric conductivity which is a measurable quantity}. {\color{black} Also, we discussed comparable calculations with ABJM theory and we suggested an experiment for our result. }

\vspace{2cm}


\end{titlepage}

\renewcommand{\thefootnote}{\arabic{footnote}}
\setcounter{footnote}{0}



\section{Introduction\label{sec:intro}}

Entanglement entropy plays a prominent role in various areas of physics such as quantum field theory, gravity theory and condensed matter physics. However, it is very difficult to evaluate this amount because of non-local property. Fortunately, an effective way to calculate entanglement entropy has recently been developed by Ryu and Takayanagi \cite{Ryu:2006bv}. Since this method is based on AdS/CFT correspondence \cite{Maldacena:1997re} which relates gravitational systems to quantum field theories, it may shed a light on understanding quantum gravity \cite{VanRaamsdonk:2009ar,VanRaamsdonk:2010pw}. Apart from this, the method is useful to study various interesting cases. For examples, it has been used to distinguish different phases {\it e.g.},\cite{Klebanov:2007ws,Albash:2012pd,Cai:2012sk} and to study thermalization under quantum quenches {\it e.g.},\cite{AbajoArrastia:2010yt,Albash:2010mv,Balasubramanian:2011ur,Balasubramanian:2010ce}. Further studies with the method show us new universal properties of generic CFT {\it e.g.},\cite{Myers:2010xs,Myers:2010tj,Bueno:2015rda,Bueno:2015xda,Park:2015dia}. Even for a particular non-conformal field theory, it turns out that Ryu and Takayanagi's proposal is still useful by considering a top-down study on the field theory and the corresponding supergravity solution \cite{Kim:2014yca,Kim:2014qpa,Kim:2016dzw,Jang:2016tbk,Jang:2016aug,Jang:2017gwd,Jang:2018dby}.

Entanglement entropy is a very important quantity but, in most cases, it is difficult to calculate. It is also unclear whether this amount can be measured easily\footnote{Recently quantum purity, R$\acute{e}$nyi entanglement entropy and mutual information was measured in experiments \cite{islam}.}. Therefore, it is hard to tell if Ryu and Takayanagi's method is valid in various cases. In this study, we try to find examples where entanglement entropy changes can be represented by other readily measurable physical quantities. Like thermal entropy, the entanglement entropy responses to other external sources. One of such sources which can be produced in a laboratory is the electric field. Thus we will investigate how the entanglement entropy changes under the electric field using holographic methods.

Accordingly, to demonstrate a system in the electric field holographically, we need to consider a gravity dual describing the system. The Ryu-Takayanagi's formula and the gravity dual allow us to study the response of entanglement entropy to the electric field. For the sake of simplicity, we introduce a time-independent electric field, i.e, DC electric field. To realize such a situation, we take the momentum relaxation into account. This is because infinite {\color{black}DC }currents can occur without the momentum relaxation. The gravity duals have been developed to study holographic DC conductivities {\color{black}in the context of AdS/CMT {\it e.g.},\cite{Horowitz:2012ky,Donos:2012js,Andrade:2013gsa,Vegh:2013sk,Horowitz:2013jaa,Blake:2013bqa,Davison:2013jba,Davison:2013txa,Donos:2013eha,Blake:2013owa,Donos:2014uba,Kim:2014bza,Gouteraux:2014hca,Blake:2014yla,Donos:2014cya,Kim:2015wba,Blake:2015ina,Seo:2015pug,Khimphun:2016ikw,Park:2016slj,Seo:2016vks} and some interesting time-dependent extensions {\it e.g.},\cite{Withers:2016lft,Davison:2016ngz,OBannon:2016exv,Bagrov:2018wzu}.} Among these backgrounds, we adopted a background geometry of an axion model, where a scalar field breaks translation invariance explicitly\cite{Andrade:2013gsa}.

In the absence of {\color{black} the bulk gauge field}, an analytic form of holographic entanglement entropy with small momentum relaxation was studied in \cite{Mozaffara:2016iwm}. In this work, we performed {\color{black} analytic and numerical studies on the strip entangling region in the presence of the bulk electric field, that is dual to the density and the external electric field in the boundary system}. {\color{black} We obtained new analytic expressions of the entanglement entropy in the very narrow strip limit. Since our study is about 2+1 dimensional field theory systems, one may compare the result to a field theoretic calculation qualitatively. So, we suggest a comparable situation in the ABJM theory \cite{Aharony:2008ug}.
Also,} we provide numerical results without taking any limit. {\color{black} Our numerical results reproduce the earlier work \cite{Gushterov:2017vnr} and extend its numerical study. This calculation shows that entanglement entropy becomes larger with more charge density and stronger momentum relaxation. We will discuss the physical implication for the extended parameter space.}

Furthermore, we studied how the strip type entanglement entropy changes by turning on the constant electric field. And we found that the minimal surface anchored to the strip is tilted by the electric field and the deformation is proportional to the thermoelectric conductivity $\bar \alpha$ in the conductivity tensor. This is a quite interesting result because the thermoelectric conductivity is a clearly measurable quantity.
However, the entanglement entropy changes from the deformation do not appear in the linear order of the electric field. To see the effect, we need to consider the quadratic order gravity dual but it is beyond our present work.

Instead we consider the wedge type entangling region whose symmetric axis is not orthogonal to the electric field
. Finally, we took a sharp wedge limit and obtained the entanglement entropy changes at the linear level of the electric field. And we confirmed that the response to the electric field is proportional to the anticipated thermoelectric conductivity $\bar \alpha$. This result is also remarkable because the thermoelectric conductivity or the Nernst signal could reflect the existence of the quantum critical point even in the normal phase of the superconductor\cite{Ong:2005,Hartnoll:2007ih,Hartnoll:2008hs,Kim:2015wba}. Thus it is suggested that the entanglement entropy may be associated with the quantum critical point.

This paper is organized as follows: In section 2 we introduce the bulk solution dual to a momentum relaxation system with the electric field and summarize how the conductivities can be related to the bulk fields. In section 3, we study on the holographic entanglement entropy for the strip entangling region. In section 4, we consider the holographic entanglement entropy for the wedge region. In section 5. we conclude our work.


\section{A Gravity Dual of Momentum Relaxation with Electric Field}

In this section we review a gravity dual to a momentum relaxation system in the electric field. The geometry has been discussed to study finite DC conductivity {\it e.g.}, \cite{Andrade:2013gsa,Donos:2014cya}. The momentum relaxation in physical systems is essential to obtain finite DC conductivity. Let us start with a simple model with momentum relaxation:
\bea\label{ActionGravity}
S = \frac{1}{16\pi G} \int d^4 x \sqrt{-g}\left( R + 6 - \frac{1}{4}F^2 -\frac{1}{2} \sum_{\mathcal I=1}^2\left(\partial \chi_{\mathcal I}\right)^2  \right)
.
\eea
This system admits a black brane solution as follows:
\bea\label{Ansatz00}
&&ds^2 = - U(r) dt^2 + r^2\left( dx^2 + dy^2 \right) + \frac{dr^2}{U(r)}~,\\
&&A=q\left(\frac{1}{r_h}-\frac{1}{r}\right)dt~,~\chi_{\mathcal I} = \left( \beta   x,  \beta  y  \right)~,
\eea
with $U(r) = \left( r^2 -\frac{\beta^2}{2}-\frac{M}{r} +  \frac{q^2}{4 r^2} \right)$\footnote{We set the AdS radius to 1 and $F = dA$.}. The massless scalar field, $\chi_{\mathcal I}$, is associated with momentum relaxation, breaking the translational symmetry and inducing finite DC conductivity. To see this, let us turn on the external electric field and other related fields along the $x$-direction
\bea		\label{def:fluctuationsfields}
\delta A_x = -E_x t + a_x(r)~~,~~
\delta g_{tx} = r^2 h_{tx}(r) ~~,~~\delta g_{rx} = r^2 h_{rx}(r)~~,~~\delta \chi_x = \psi_x(r)~~,
\eea
where $E_x$ corresponds to the external field and other fluctuations are dual to the electric current and the heat current in the boundary system.

Then, the linearized equations of motion are given as follows :
\bea
&&h_{t x}''(r)+\frac{4 \,h_{t x}'(r)}{r}-\frac{\beta ^2 h_{t x}(r)}{r^2\, U(r)}+\frac{q \,a_x'(r)}{r^4} = 0
~~,~~\psi _x'(r)-\beta  h_{r x}(r)-\frac{q E_x}{\beta\,  r^2 U(r)}=0~,\label{EinsteinFluc}
\\
&&a_x''(r)+\frac{U'(r) a_x'(r)}{U(r)}+\frac{q h_{t x}'(r)}{U(r)}=0~,\label{MaxwellFluc}
\\
&&\psi _x''(r)+\left(\frac{U'(r)}{U(r)}+\frac{2}{r}\right) \psi _x'(r)-\beta h_{rx}'(r)-\beta\left(\frac{ U'(r)}{U(r)}+\frac{2}{r}\right)h_{rx}(r)=0\label{AxionFluc}~~,
\eea
where Eq. (\ref{EinsteinFluc}) comes from the Einstein equation, and Eq. (\ref{MaxwellFluc}) and (\ref{AxionFluc}) are the equations of motion of the matter fields. Note that all the equations above are not independent because Eq. (\ref{AxionFluc}) can be obtained by rearranging the other equations. As a consequence, only three equations in (\ref{EinsteinFluc}) and (\ref{MaxwellFluc}) determine the profile of the fluctuations for given boundary conditions. Another thing we need to know is that $h_{rx}$ is not a dynamical field and it plays a role of a Lagrange multiplier. In fact, this is a shift vector in an ADM decomposition along the holographic radial direction. We will discuss the gauge fixing later.

The electric current and the heat current using gauge/gravity duality are given by
\bea
&J^x = \lim_{r\to\infty} \mathcal{J}(r) ~~,~~
Q^x = T^{tx}-\mu J^x = \lim_{r \to \infty} \mathcal{Q}(r)~~,
\eea
where
\bea
\mathcal{J}(x)&\equiv&  \sqrt{-g} F^{xr} =U(r)~a_x'(r)+q~h_{tx}(r) ,
\nonumber
\\
Q(r)&\equiv& U^2(r) \left( \frac{\delta g_{tx}}{U(r)} \right)'- A_t(r)\mathcal{J}(r)~~.
\eea
In addition, it can be easily checked that $\mathcal J(r)$ and $\mathcal Q(r)$ remain as constants along the holographic radial direction $r$ by using the equations of motion (\ref{EinsteinFluc}-\ref{AxionFluc}). Therefore, the boundary currents can be computed at the horizon $r=r_h$
\bea
&J^x = \lim_{r\to r_h} \mathcal{J}(r) ~~,~~
Q^x = \lim_{r \to r_h} \mathcal{Q}(r)~~,
\eea
Near the horizon the fields behave as
\bea
&&a_x\sim -\frac{ E_x \log(r-r_h)}{4 \pi T} + a_{x}^{(0)}+ \mathcal{O}\left(r-r_h\right)~~,~~h_{tx}\sim  h_{tx}^{(0)}+ \mathcal{O}\left(r-r_h\right)~~,
\nonumber
\\
&&h_{rx}\sim \frac{\mathbb{H}_{rx}}{r^2 U(r)} + h_{rx}^{(0)} + \mathcal{O}\left(r-r_h\right)~~,~~\psi_x \sim  \psi_{x}^{(0)}+ \mathcal{O}\left(r-r_h\right)~~,
\nonumber
\\
&&U(r)\sim4\pi T(r-r_h)+\cdots,
\eea
where the Hawking temperature $T$ is given by $T= \frac{U'(r_h)}{4\pi}$ and the logarithmic term appears due to the ingoing boundary condition. By solving the equations of motion near the horizon, one can obtain
\bea
{\mathbb{H}_{rx} =  r_h^2 h_{tx}^{(0)}}~~,~~h_{tx}^{(0)} =  - \frac{q E_x}{\beta^2 r_h^2}~~.
\eea
Using these results, the linear response theory yields the following electric conductivity ($\sigma=J^x/E_x$) and thermoelectric conductivity ($\bar{\alpha}=Q^x/E_x$):
\bea
\sigma = 1-\frac{q h_{tx}^{(0)}}{E_x} = 1+ \frac{\mu^2}{\beta^2}~~~,~~~\bar\alpha = - \frac{4\pi r_h^2 h_{tx}^{(0)}}{E_x} = \frac{4\pi q}{\beta^2} .
\eea
So the constant $\mathbb{H}_{rx}$ is given by a combination of physical parameters as $\mathbb{H}_{rx} = - \frac{1}{4\pi}\bar\alpha E_x $~.

The following sections will consider minimum surfaces in this geometry. It is convenient to use a formal expansion parameter $ \lambda $, which will be taken as 1 after the calculation. This trick allows us to write the metric in the following form:
\bea
ds^2&=&-\,U(r)dt^2+\frac{dr^2}{U(r)}+r^2(dx^2+dy^2)\label{metric}\\&&+2\,\lambda\biggl(
\delta g_{tx}(r)dtdx\,+\delta g_{rx}(r)drdx
\biggr)+ \mathcal{O}\left(\lambda^2\right)~.\nonumber
\eea
In addition to the conductivities, other physical quantities such as temperature, energy density, charge density, and entropy density can be read from the geometry as follows:
\bea
&&T = \frac{U'(r_h)}{4\pi}=\frac{3r_h}{4\pi}\left(
1-\frac{\beta^2}{6r_h^2}-\frac{q^2}{12r_h^4}\right)~,~\\&&\epsilon = 2M=\frac{4r_h^4-2r_h^2\beta^2+q^2}{2\,r_h}~,~\rho =  q~,s = 4\pi r_h^2~.
\eea
These quantities characterize the dual system to the unperturbed geometry with $\lambda=0$.

\section{Holographic Entanglement Entropy in Electric Field: Strip Entangling Region \label{sec:sec2}}

\subsection{Basic Setup}

In this subsection, we use the Ryu-Takayanagi formula to obtain entanglement entropy for the strip region. The strip also extends along a direction perpendicular to the electric field. To describe the minimal surface anchored to the strip, we may consider a two dimensional surface with the following map in the target geometry (\ref{metric}):
\begin{eqnarray}\label{mapping}
t = 0 ~~,~~ r= r(\sigma_1) ~~,~~x=\sigma_1~~,~~y = \sigma_2~~,
\end{eqnarray}
where $y$ coordinate indicates the infinitely extending direction. Then, one can find the action for the minimal surface as follows:
\bea 					\label{eq:HEEperpendicular0}
\mathcal{A} = \int_{-L/2}^{L/2} dy \int_{-l/2}^{l/2} dx \sqrt{ \frac{r^2 r'^2}{U(r)} + r^4 + 2 \lambda r^2 \delta g_{rx} r'}~~,
\eea
where $L$ is the length of the strip along $y$-direction. This is our basic action for the holographic entanglement entropy. Or, if we define another convenient coordinates $z \equiv \frac{1}{l r}$ and $\sigma \equiv x/l$, we can write down the action in terms of $z(\sigma)$ as follows:
\begin{eqnarray}\label{SLz}
\mathcal{A} = \frac{L}{l}\int_{-1/2}^{1/2} d\sigma \sqrt{ \frac{  z'^2}{ z^6  f(z)} + \frac{1}{z^4} - 2 \lambda\frac{z'}{z^4} \delta g_{rx}(z) }~~,
\end{eqnarray}
where  $f(z) = \frac{1}{z^2} -\frac{{\tilde\beta}^2}{2}- \tilde{M} {z} + \frac{\tilde{q}^2}{4} z^2 $ with $\tilde \beta = \beta l$, $\tilde M = M l^3$ and $\tilde q = q l^2$. $r(x)$ can always be recovered by $r(x)=\frac{1}{ l z (x/l)}$.

The Lagrangian in (\ref{SLz}) has no explicit dependence of $\sigma$. So there is a conserved quantity $H$ given by
\begin{eqnarray}
H = \frac{L}{l} \frac{1-\lambda \delta g_{rx} z'}{z\sqrt{ \frac{z'^2}{f(z)}+ z^2( 1 -2\lambda z' \delta g_{rx}})}=\frac{L}{l z_*^2}~~,
\end{eqnarray}
where $z_* \equiv z(\sigma_*)$ is the location of the tip of the minimal surface. Since we are considering regular surfaces, $z'(\sigma)$ vanishes at the tip of the surface. From this relation one can find the integrated first order equations of motion as follows :
\begin{eqnarray}\label{eomFirstOrder}
z'&=& \pm  \frac{\sqrt{f(z)\left(z_*^4-z^4\right) }}{z} - \lambda \,\, \frac{f(z) \delta g_{rx}(z)\left(z_*^4-z^4\right)}{z^2} + \mathcal{O}\left( \lambda^2 \right)\\
&=& \pm \frac{\sqrt{f(z)\left(z_*^4-z^4\right) }}{z} +  \lambda \,\,  E_x\frac{\bar \alpha l^4 \left(z_*^4-z^4\right)}{4\pi}+ \mathcal{O}\left( \lambda^2 \right)~~,\nonumber
\end{eqnarray}
where we took a gauge $\delta g_{rx} = \frac{\mathbb H_{rx}}{r^2 U(r)}$, which is consistent with the boundary conditions\footnote{In the calculation we took a gauge $h_{rx} = \frac{\mathbb{H}_{rx}}{r^2 U(r)}$. It is legitimate because $h_{rx}$ plays a role of a Lagrange multiplier at the linear level and the choice satisfies the regularity condition at the horizon and near the boundary of AdS space. Furthermore, our final result in this work depends on the gauge invariant quantities, such as the thermoelectric conductivity $\bar\alpha$ and geometric angle $\delta$. For this reason, the gauge fixing is believed to be adequate to consider this calculation. In addition, we speculate that the gauge fixing would be clearer if we consider the second order effect of the electric field on the geometry. This is because  additional constraints must be considered for $g_{rx}$.}. Plugging the above equations into (\ref{SLz}), we obtained the regularized area for the minimal surface as follows:
\begin{eqnarray}\label{Areg}
\mathcal{A}_{reg}^{\text{strip}} =\frac{2L}{l} \int_{\epsilon/l}^{z_*} dz \frac{z_*^2}{z^3\sqrt{f(z)\left(z_*^4-z^4\right)}}~~,
\end{eqnarray}
where $\epsilon$ is the length scale corresponding to the UV cutoff. Here, one can notice that the effect from the electric field is encoded in the location of the tip $z_*$. By using the Ryu-Takayanagi formula, the entanglement entropy for the strip is:
\begin{eqnarray}
S_{EE}^{\text{strip}}= \frac{\mathcal{A}^{\text{strip}}_{reg}}{4\,G_N}~~,
\end{eqnarray}
where $G_N$ is the 4-dimensional Newton constant.
In order to show the numerical result, we define the following refined function which is $\frac{4\,G_N l}{L}$ times the finite part of the entanglement entropy up to the minus sign:
\begin{eqnarray}\label{RHEE}
\hat{S}_{R}^{\text{strip}} \equiv  \frac{2 l}{\epsilon}-\frac{4\, G_N l}{L} S^{\text{strip}}_{EE}~~.
\end{eqnarray}

\begin{figure}
\centering
\includegraphics[width=0.45\textwidth]{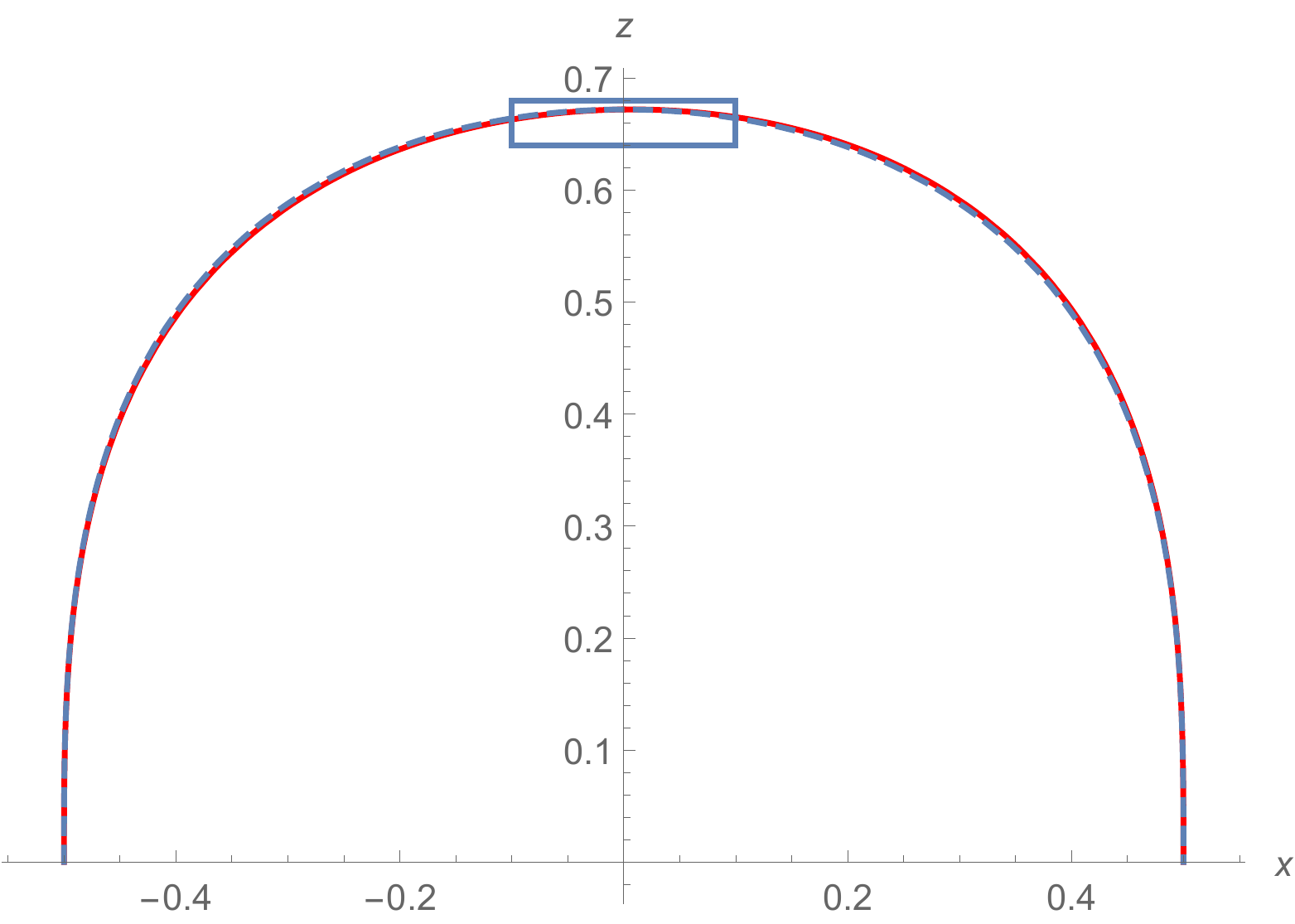}
\includegraphics[width=0.45\textwidth]{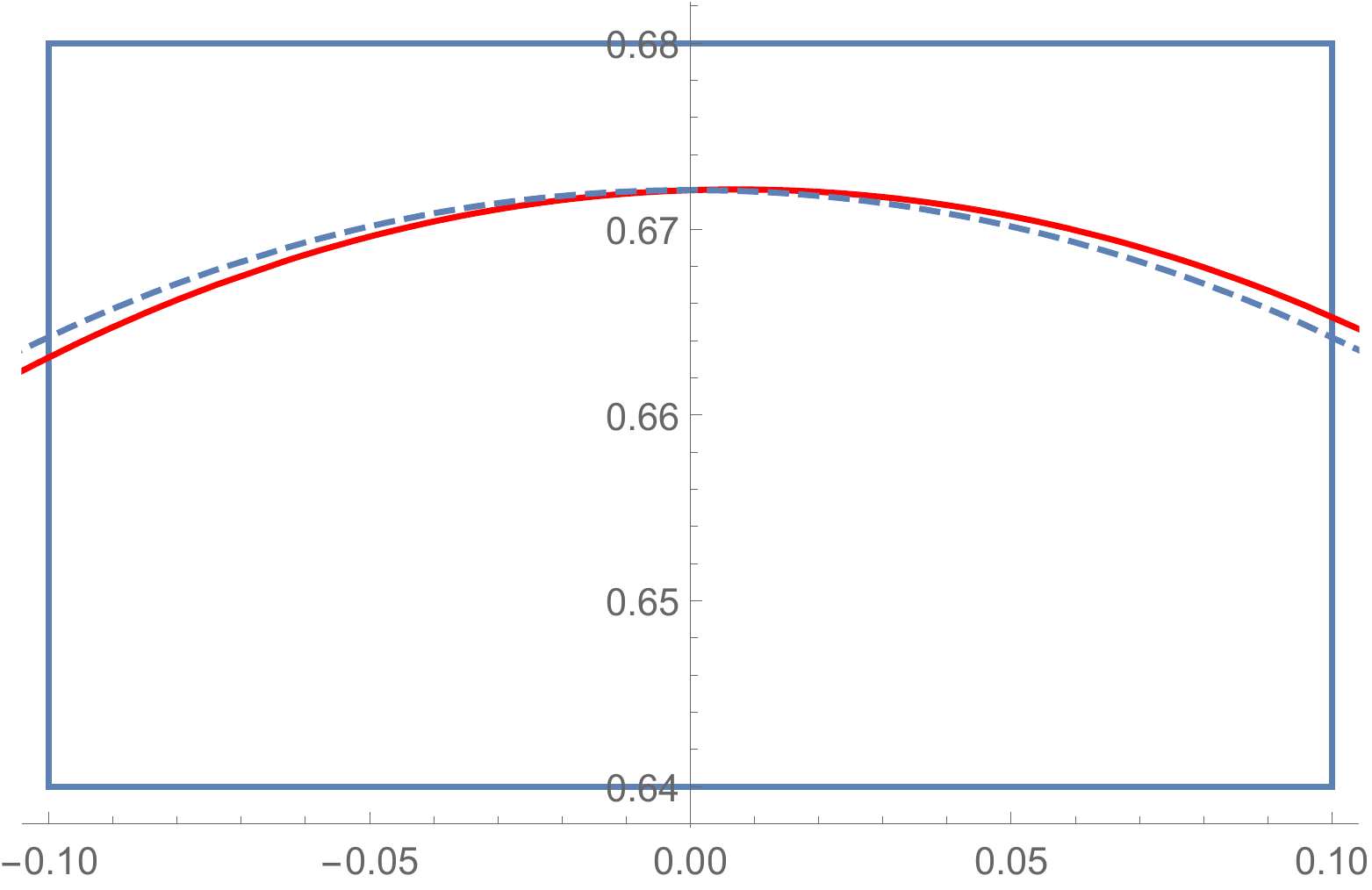}
\caption{\label{fig1} The intersection of a deformed minimal surface by the electric field (Solid curve) with $\tilde\beta=1$, $\tilde M=1$ and $\tilde q=1$: The dotted line denotes the minimal surface before applying the electric field. {\color{black}$\lambda = 0.05$ is taken for visualization and the right figure is the enlarged picture for the blue square near the tip.}}
\end{figure}

Now, let us look at the equation of motion for the minimal surface. The equation of motion can be derived from the action (\ref{SLz}) and the solution is expanded in terms of $\lambda$ as follows :
\begin{eqnarray}
z(\sigma) = z_0(\sigma) + \lambda z_1(\sigma)+ \mathcal{O} \left( \lambda^2 \right)~.
\end{eqnarray}
Then the equation of motion for each order is given by
\begin{eqnarray}
&&z_0''-\frac{z_0'{}^2 f'\left(z_0\right)}{2 f\left(z_0\right)}+2 z_0 f\left(z_0\right)+\frac{z_0'{}^2}{z_0}=0~,\label{zerothEq}\\
&&z_1''+\frac{z_1 f'\left(z_0\right){}^2 z_0'{}^2}{2 f\left(z_0\right){}^2}+2 z_0 z_1 f'\left(z_0\right)+\frac{z_0' \left(-z_1 f''\left(z_0\right) z_0'+\delta g_{rx}\left(z_0\right) f'\left(z_0\right) z_0'{}^2-2 f'\left(z_0\right) z_1'\right)}{2 f\left(z_0\right)}\nonumber\\&&~~~~+2 f\left(z_0\right) \left(z_1-3 z_0 \delta g_{rx}\left(z_0\right) z_0'\right)+\delta g_{rx}'\left(z_0\right) z_0'{}^3+\frac{2 z_0' z_1'-\delta g_{rx}\left(z_0\right) z_0'{}^3}{z_0}-\frac{z_1 z_0'{}^2}{z_0^2}=0~.\label{firstEq}
\end{eqnarray}
Here, the zeroth order solution is given by an even function from the symmetry of (\ref{zerothEq}). Furthermore, suppose $z_1(\sigma)$ is a solution of (\ref{firstEq}), then one can easily show that $-z_1(-\sigma)$ is also a solution. Therefore the solution $z_1$ of (\ref{firstEq}) is an odd function. By using this fact, one can recognize that $z_*$ is larger than $z_0(0)$ because $z_0'(0)$ vanishes but there is a non-vanishing $z_1'(0)$. We found a solution, $z_0(\sigma) + \lambda z_1(\sigma)$, numerically and plot the solution in Fig. \ref{fig1}.

To find the effect on entanglement entropy, let us come back to (\ref{Areg}). In order to get the $z_*$, one can consider a shift of the tip along the $\sigma$ direction. It is denoted by $\lambda \Delta$, then the regularity condition of the tip is given by
\begin{eqnarray}
z_0'(\lambda \Delta) + \lambda z_1'(\lambda \Delta) = 0~.
\end{eqnarray}
This condition gives us $\Delta = - z_1'(0) / z_0''(0)$. Therefore, it turns out that $z_*$ is
\begin{eqnarray}
z_* \sim z_0(\lambda \Delta) + \lambda z_1(\lambda \Delta) &=& z_0(0) +\frac{1}{2}\lambda^2 \Delta^2 z_0''(0)+\lambda^2 \Delta z_1'(0) + \mathcal O \left(\lambda^3\right)\nonumber\\&=&z_0(0)-\frac{\lambda ^2 z_1'(0)^2}{2 z_0''(0)}+ \mathcal O \left(\lambda^3\right)~~.
\end{eqnarray}
From this, we found that the tip change is the order of $\lambda^2$. However, our background (\ref{metric}), is only valid in the linear order. Therefore, one needs a quadratic background metric to see the effect on entanglement entropy for the strip case. We leave this as a future work.

{\color{black} Although this result doesn't give us the linear variation of entanglement entropy, it is still worth considering the zeroth order calculation as a leading part of another specific kind of entangling region. We will continue our study to a case with the wedge type entangling region in the section \ref{Section:wedge}. Such a region is regarded as a tail part of the strip case. See the cartoon in Fig. \ref{fig2}. Also, once the very sharp wedge limit is taken ($\Omega \to 0$), this limit should give rise to the strip result. Due to these reasons, it is desirable to scrutinize the zeroth order calculation for the strip case numerically and analytically in the following subsections.}

\subsection{Numerical Result for Holographic Entanglement Entropy}

Now we study the holographic entanglement entropy (\ref{Areg}) numerically. As examined in the previous subsection, there is no linear variation of the entanglement entropy to the electric field \textcolor{black}{for the strip entangling region}. Therefore, \textcolor{black}{in this subsection}, we will focus only on \textcolor{black}{numerical results of the entanglement entropy} without the electric field .

\begin{figure}
\centering
\begin{subfigure}[]
\centering
\includegraphics[width=0.3\textwidth]{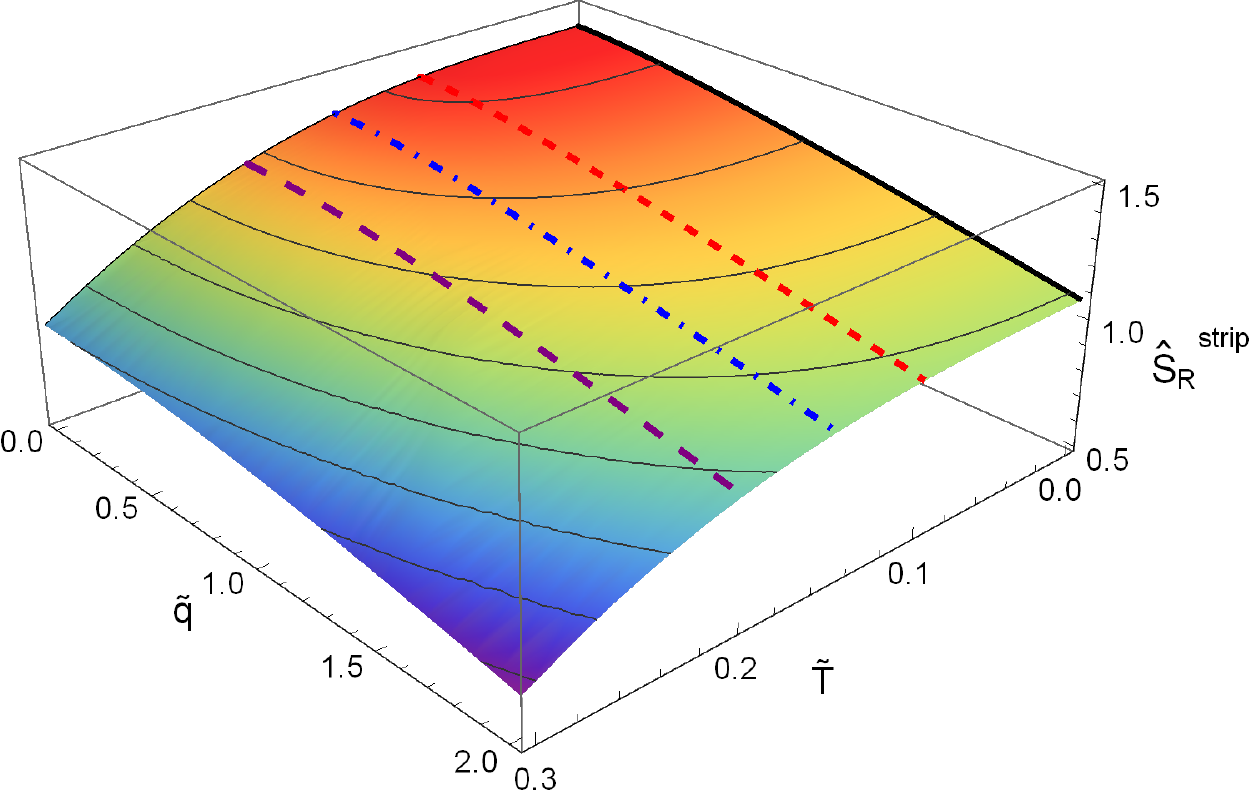}
\end{subfigure}
\begin{subfigure}[]
\centering
\includegraphics[width=0.3\textwidth]{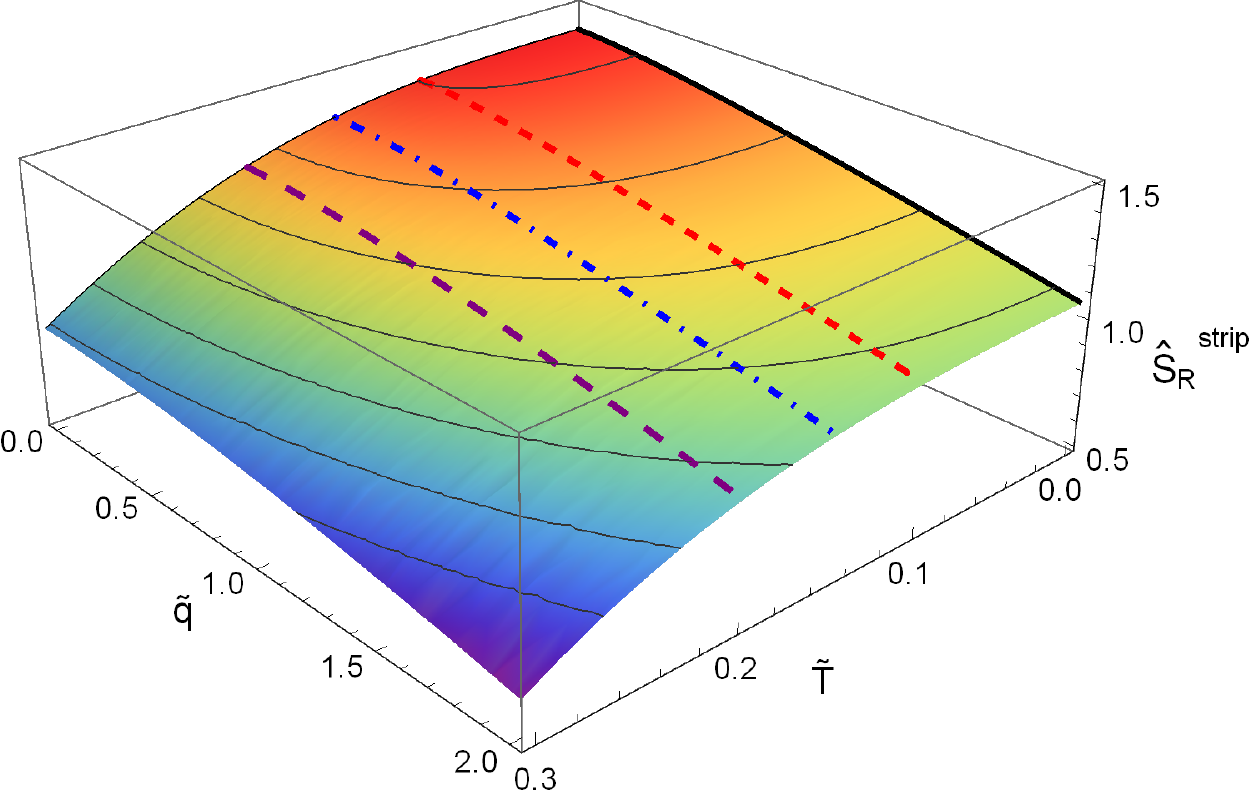}
\end{subfigure}
\begin{subfigure}[]
\centering
\includegraphics[width=0.3\textwidth]{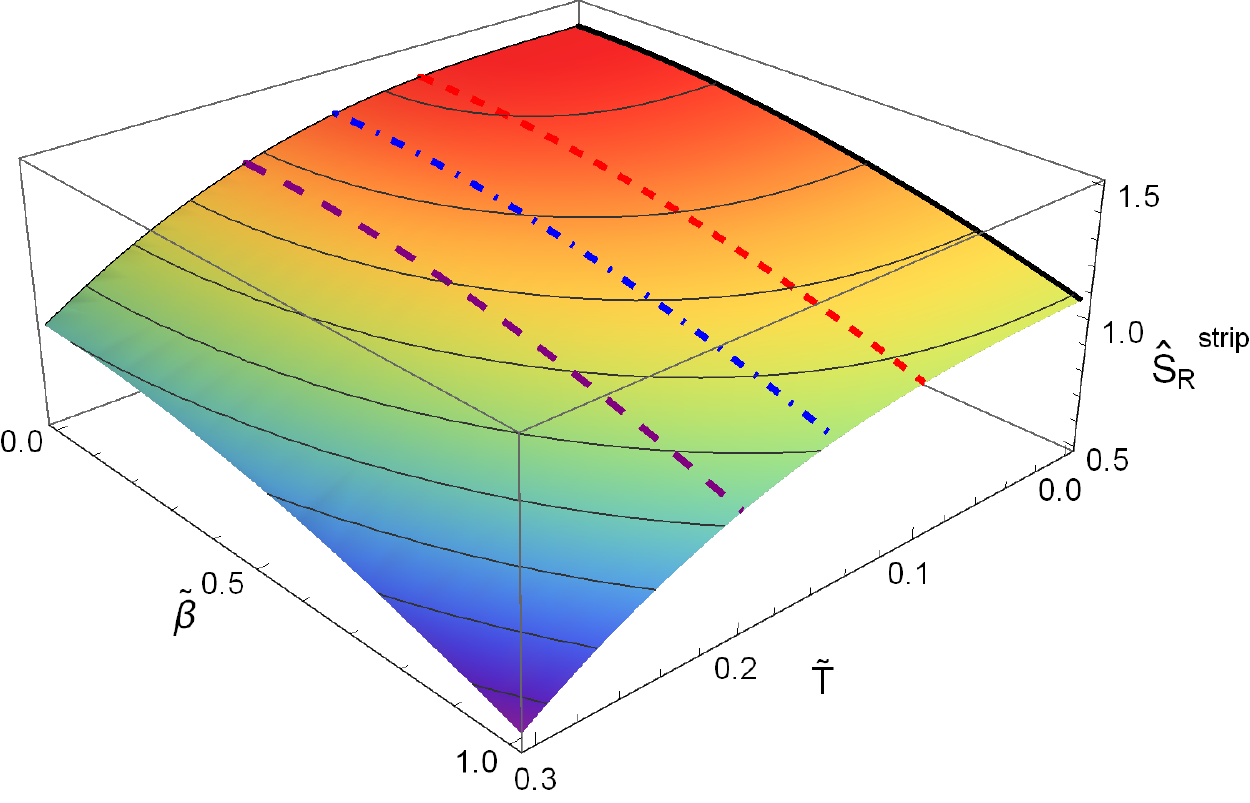}
\end{subfigure}
\begin{subfigure}[]
\centering
\includegraphics[width=0.3\textwidth]{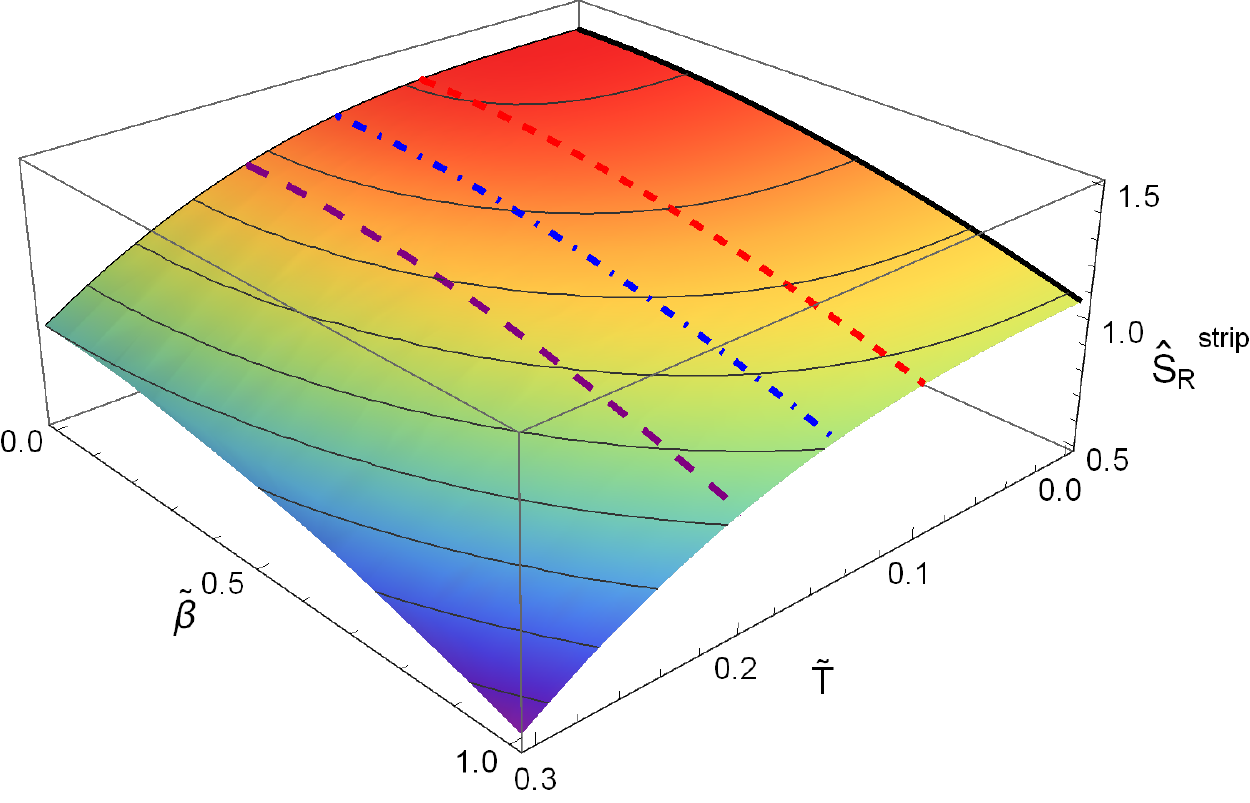}
\end{subfigure}
\begin{subfigure}[]
\centering
\includegraphics[width=0.3\textwidth]{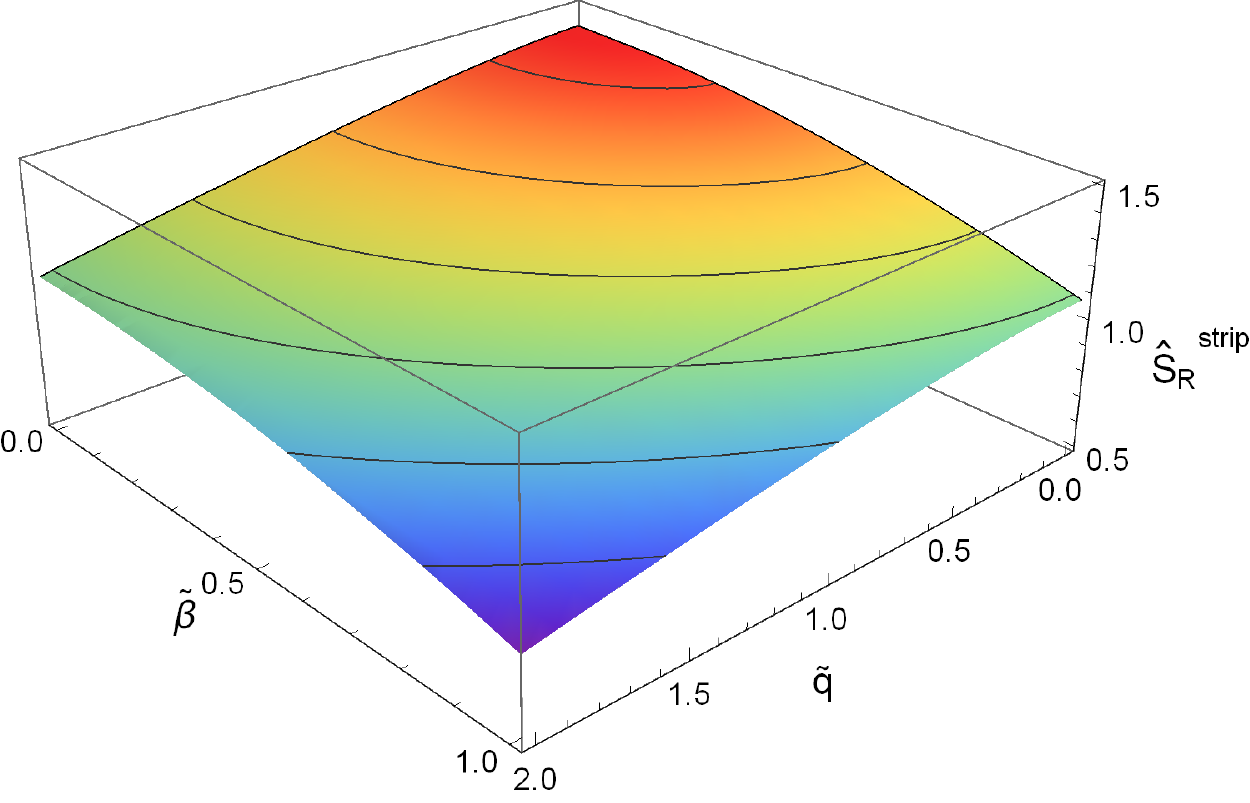}
\end{subfigure}
\begin{subfigure}[]
\centering
\includegraphics[width=0.3\textwidth]{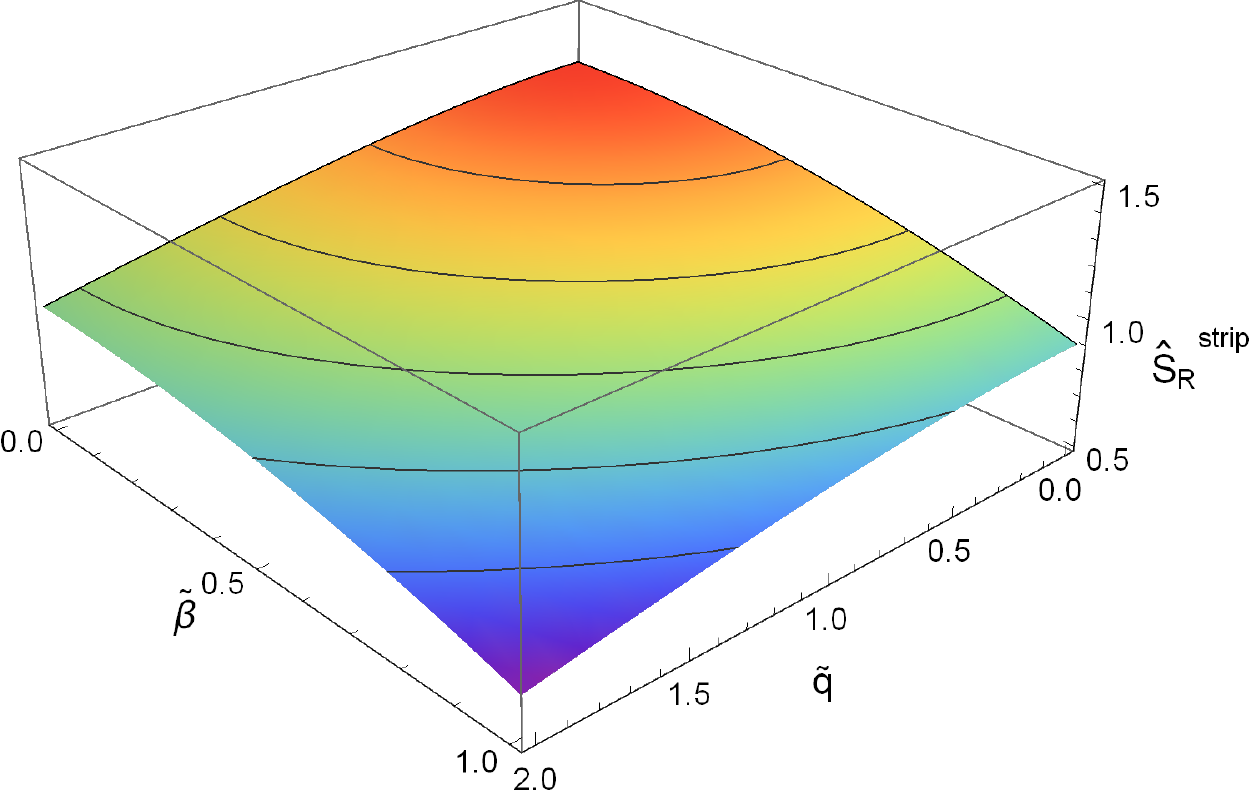}
\end{subfigure}
 \caption{The refined function $\hat S_R^{\text{strip}}$ at (a)~$\tilde{\b}=0$\,, (b)~$\tilde{\b}=0.2$\,, (c)~ $\tilde{q}=0$\,, (d)~$\tilde{q}=0.2$\,, (e)~$\tilde{T}=0$\, and (f)~$\tilde{T}=0.2$\, : The lines on the surfaces denote $\tilde T$ = 0\,(Solid), 0.1\,(Shortdashed), 0.15\,(Dotdashed), 0.2\,(Longdashed) and these lines are plotted in Fig. \ref{fig2d}. This result is easily extended to the entanglement entropy in the presence of magnetic field. Thus one may regard $\tilde q$ and $\sqrt{\tilde q^2 + \tilde B^2}$, where $\tilde B = F_{xy} l^2$}\label{fig3d}
\end{figure}

\begin{figure}[h]
\centering
\begin{subfigure}[]
\centering
\includegraphics[width=0.35\textwidth]{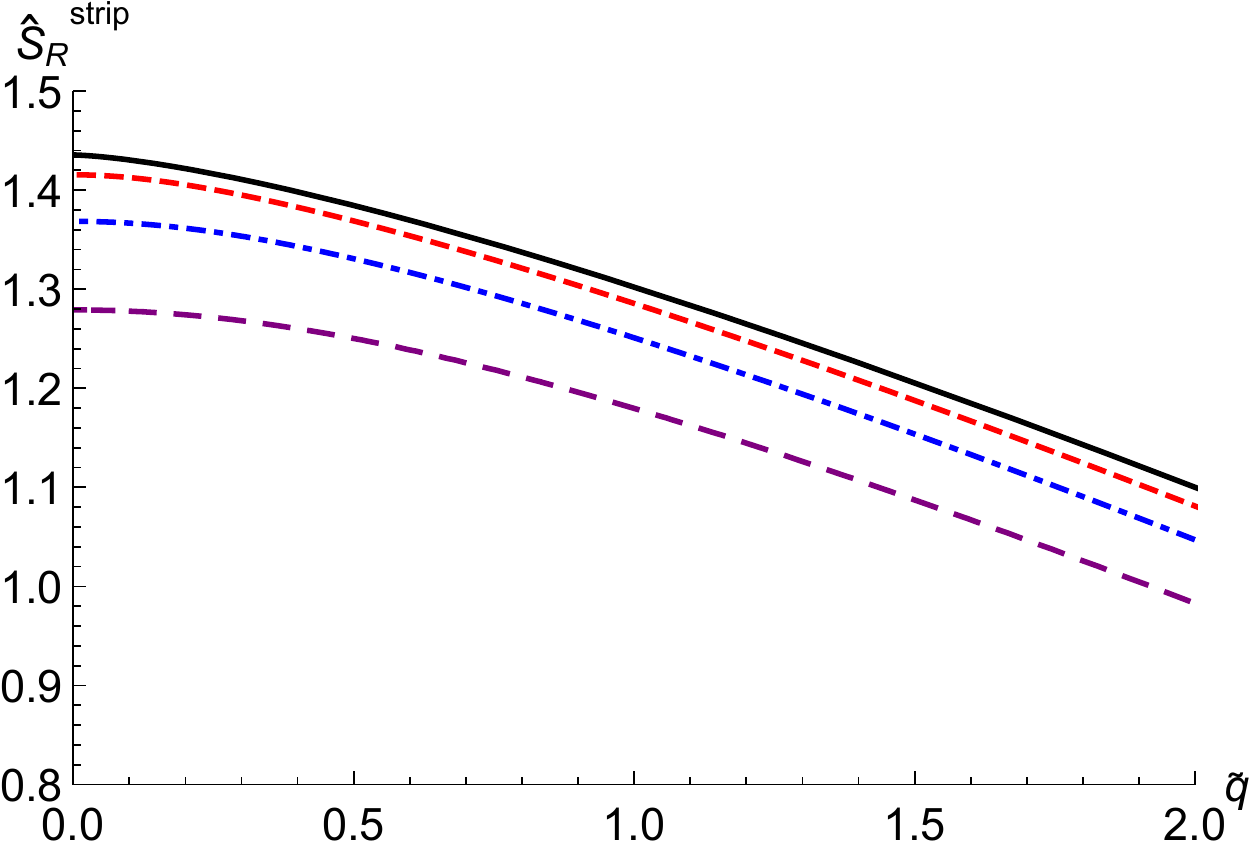}
\end{subfigure}\qquad
\begin{subfigure}[]
\centering
\includegraphics[width=0.35\textwidth]{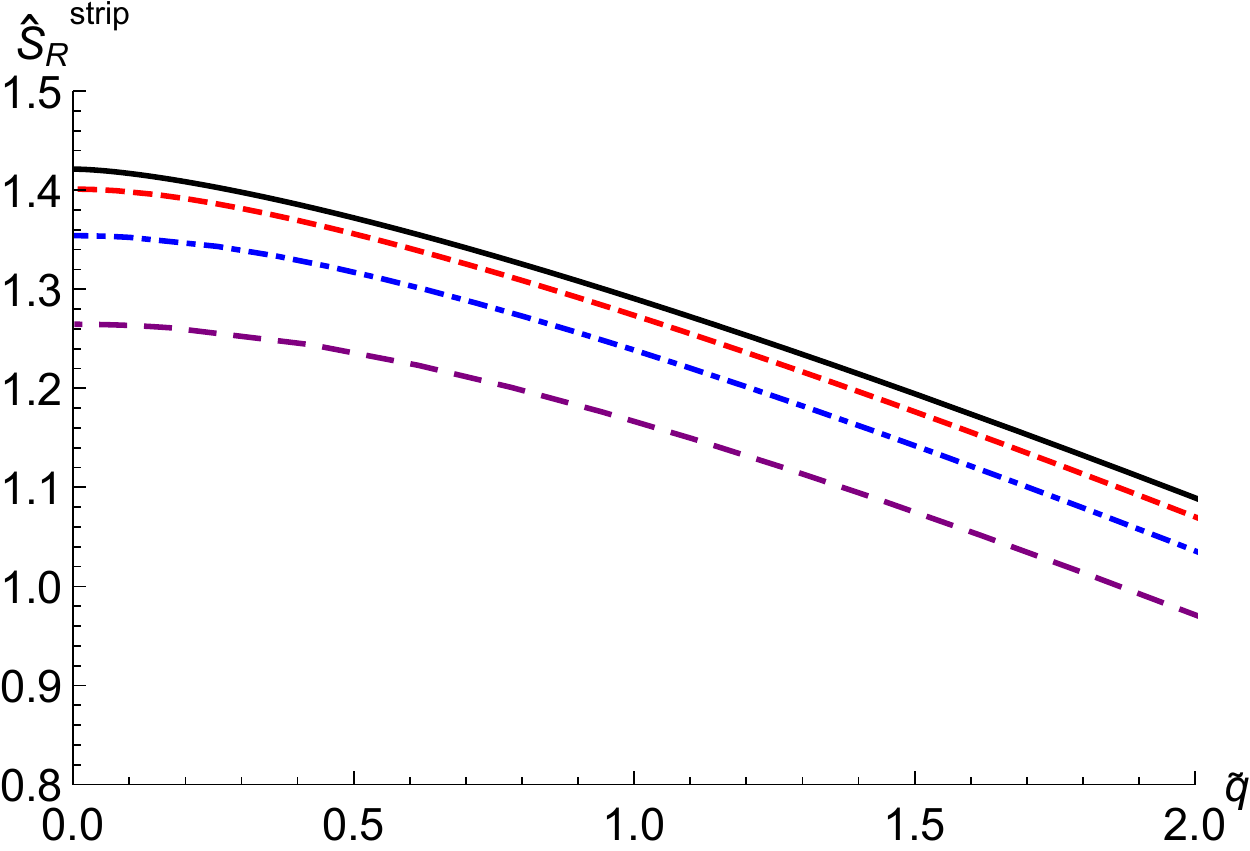}
\end{subfigure}
\begin{subfigure}[]
\centering
\includegraphics[width=0.35\textwidth]{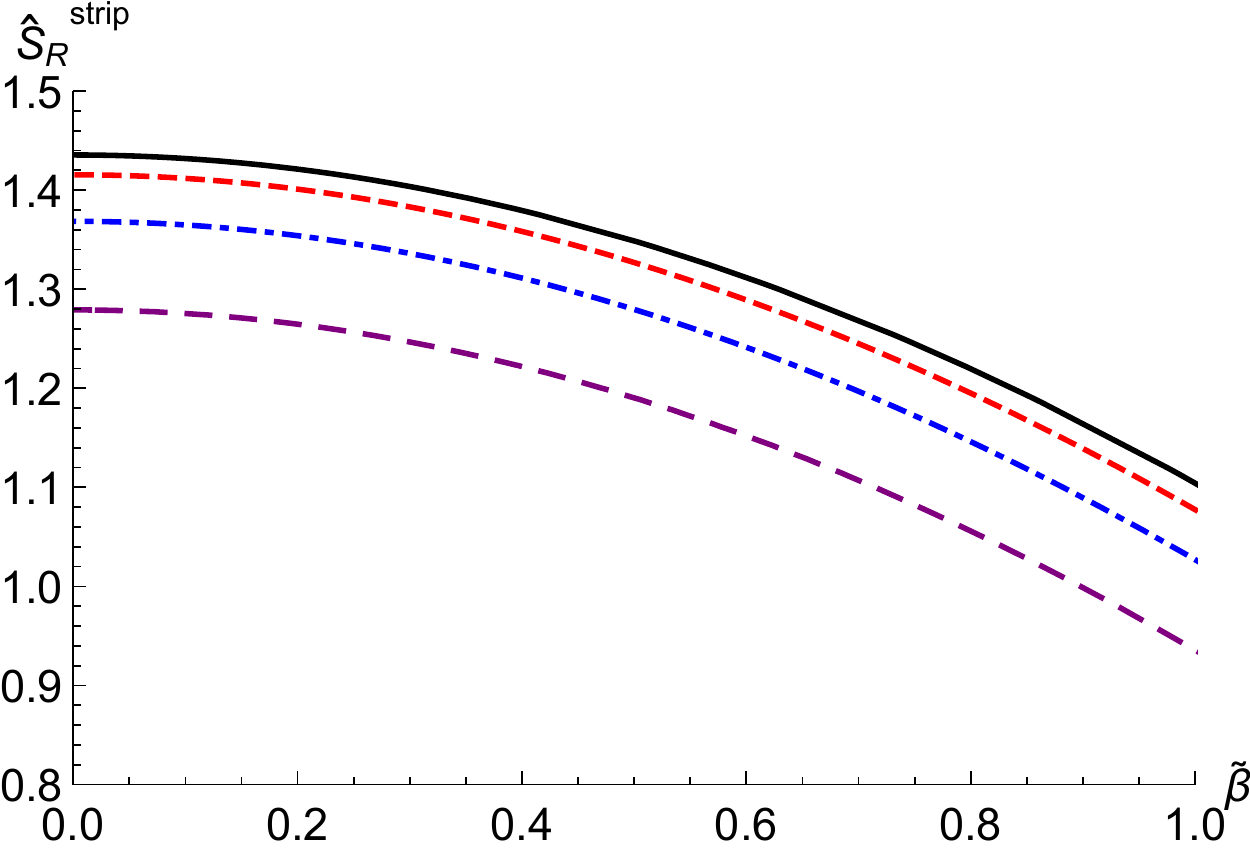}
\end{subfigure}\qquad
\begin{subfigure}[]
\centering
\includegraphics[width=0.35\textwidth]{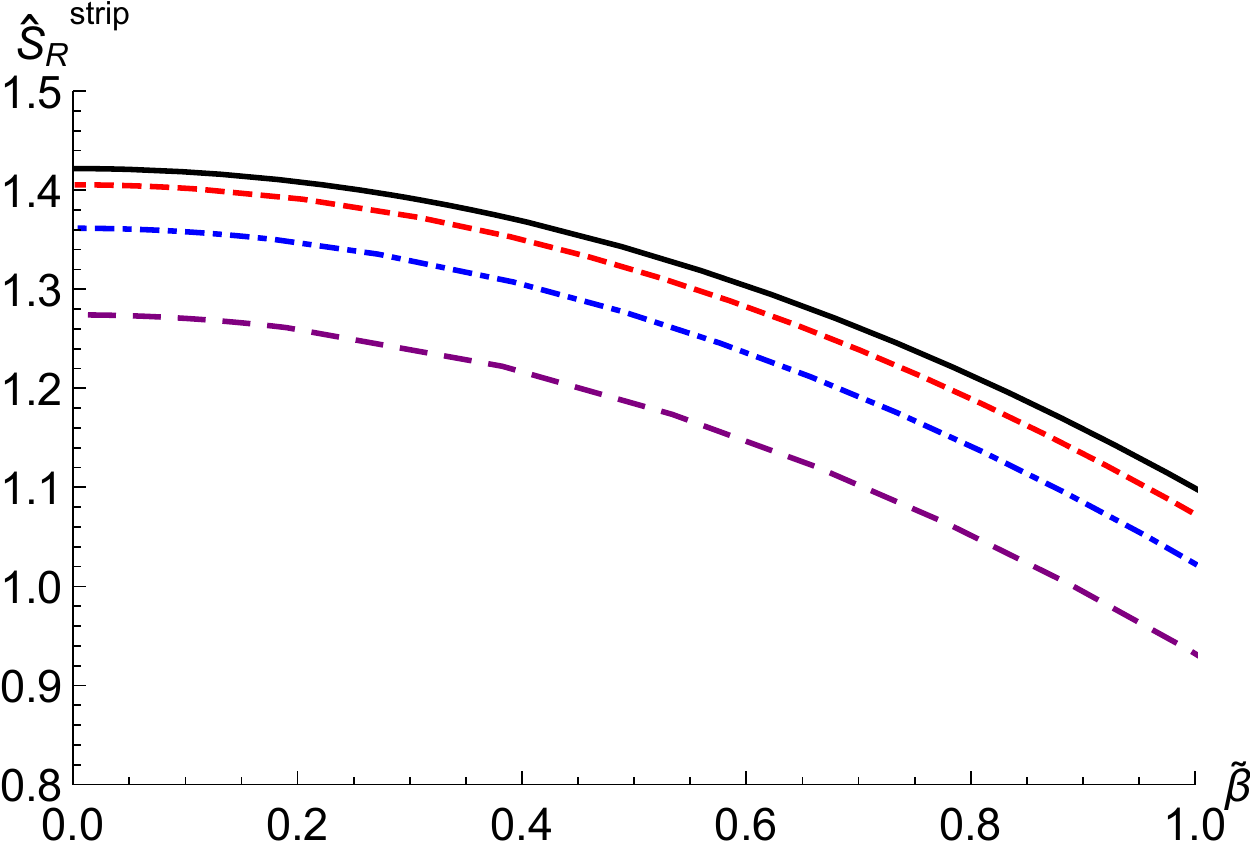}
\end{subfigure}
\caption{The refined function $\hat S_R^{\text{strip}}$
for (a) $\tilde{\b}=0$,\,(b) $\tilde{\b}=0.2$,\,(c) $\tilde{q}=0$ and (d) $\tilde{q}=0.2$. These curves represent the lines on the surfaces in Fig.\,\ref{fig3d} and correspond to the temperatures $\tilde{T}=0$(Solid), 0.1(Shortdashed), 0.15(Dotdashed) and 0.2(Longdashed).}\label{fig2d}
\end{figure}

First, we define the following function to facilitate numerical computation:
\begin{eqnarray}\label{Areg0}
\tilde{A}_0 \equiv A^{(0)}_{reg}\ell = 2 L \int_{\tilde{\e}}^{z_0^*} dz \frac{{z_0^*}^2}{z^3\sqrt{f(z)\left({z_0^*}^4-z^4\right)}}~,
\end{eqnarray}
where ${z_0^*}\equiv z_0(0)$ and $\tilde{\e}\equiv \e / l$. For convenience, we replace $\tilde{M}$ with $z_h$, \textcolor{black}{$\tilde{\b}$  and $\tilde{q}$} using the relation $f(z_h)=0$. Then, the useful expression for $f(z)$ is given by
\begin{eqnarray}
f(z)=\frac{\left(z-z_h\right) \left( z^3 \tilde{q}^2 z_h^3  + 2 z^2  \tilde{\beta }^2 z_h^2  -4 z_h^2-4 z z_h-4 z^2\right)}{4 z^2 z_h^3}.
\end{eqnarray}
By adopting a scaled coordinate $u\equiv z/z_0^*$, \textcolor{black}{we reparameterize  $\xi\equiv z_0^*/z_h$, $\bar q \equiv \tilde q z_h^2$ and $\bar \beta = \tilde \beta z_h$.}\footnote{Thus the parameters in terms of the original physical quantities are given by $\xi=\frac{r_h}{r_0^*}$, $z_h=1/(r_h l)$, $\bar \beta = \beta/r_h$ and $\bar q = q /r_h^2 $.} Then the area function (\ref{Areg0}) becomes
\begin{eqnarray}\label{numericalEE}
\tilde{A}_0 = \frac{2 L}{z_0^*}\int_{\frac{\tilde{\epsilon} }{z_0^*}}^{1} du \frac{1}{u^2 \sqrt{\left(1-u^4\right)\left(1-\xi  u\right)\left(1+ \xi u+ \xi ^2 u^2 -\frac{1}{2} \bar{\beta }^2 \xi ^2 u^2 - \frac{1}{4}\bar{q}^2 \xi ^3 u^3 \right)}}~.
\end{eqnarray}
Also, the temperature of the system can be rewritten in terms of the dimensionless parameters as
\begin{eqnarray}
 \tilde{T} \equiv T \ell = \frac{1}{16 \pi z_h}\left(12-\tilde{q}^2 z_h^4-2 \tilde{\beta }^2 z_h^2\right)
 =\frac{\xi}{16\pi z_0^*}\left( 12 - \bar q^2 - 2 \bar \beta^2  \right)~.
\end{eqnarray}
In addition, (\ref{eomFirstOrder}) and the regularity determines $z_0^*$ as follows:
\begin{eqnarray}\label{numericalzstar}
\frac{1}{z_0^*} = \int_0^1 du \frac{2 u^2}{\sqrt{\left(1-u^4\right)\left(1-\xi  u\right)\left(1+ \xi u+ \xi ^2 u^2 -\frac{1}{2} \bar{\beta }^2 \xi ^2 u^2 - \frac{1}{4}\bar{q}^2 \xi ^3 u^3 \right)}}.
\end{eqnarray}
Therefore, for given $\xi, \bar{q}$ and $\bar{\b}$, the regularized area of minimal surface is determined in terms of $\tilde{\e}$. In general, this area has a leading term that is proportional to the inverse of $ \tilde \epsilon$. Subtracting this term and taking the suitable rescaling by the definition in (\ref{RHEE}), we can get the finite part of the minimal surface area that does not depend on $ \tilde\epsilon $. We show our numerical results \textcolor{black}{in various parameter regions} in Fig.\,\ref{fig3d} and Fig.\,\ref{fig2d}. \textcolor{black}{In {\cite{Gushterov:2017vnr}}, the authors studied the system with momentum relaxation and showed their results in some parameter regions. Our results not only reproduce theirs well but also can be expanded to other parameter regions. }

{\color{black} Our result for the extended parameter space is telling us the role of impurity density, charge density and temperature in the entanglement entropy. As one can see, these parameters make the finite part of the entropy decrease. This finite part actually follows not an area law but a volume law. This decreasing behavior is coming from the F-theorem\footnote{\color{black}For the correct argument, we have to use the renormalized entanglement entropy. Its prescription can be found in \cite{Liu:2012eea}. In fact we check that the decreasing behavior is similar for a small size of entangling region covered by Fig. \ref{fig3d}.   }\cite{Casini:2012ei} due to a marginal or relevant deformation dual to the massless scalars or the bulk electric field, respectively. See \cite{Nishioka:2018khk} for a review. Also, the entanglement entropy can be affected by the temperature. As the temperature increases, thermal fluctuations is growing. These thermal fluctuations may wash out the entanglement between the entangling region and its complement. Thus one can naturally deduce that the entanglement entropy decreases with increasing of the temperature.

It is noticeable that the above zeroth order calculation is invariant under the electromagnetic duality $(F \leftrightarrow \tilde F)$ and also the rotation of $F$ and $\tilde F$ in the bulk. This property leads to the symmetry exchanging and rotating the charge density and the external magnetic field in the dual field theory \cite{Hartnoll:2007ip,Kim:2015wba}. Thus the form of entanglement entropy can be extended to the entanglement entropy in the presence of the external magnetic field. The extended entanglement entropy is simply obtained by a transform : $\tilde q^2 \to \tilde q^2 + \tilde B^2$, where $\tilde B = B l^2$ and $B$ is nothing but the magnetic field $F_{xy}$ of the black brane. Therefore our result can cover the dyonic black brane case by a simple parameter change.


 }

\subsection{Analytic Result for Holographic Entanglement Entropy}
\label{section3}

Now we try to obtain analytic expression for the holographic entanglement entropy in {\color{black}the small $l$} limit. Our formulas (\ref{numericalEE}) and (\ref{numericalzstar}) are also useful to study the analytic form. In order to study the analytic expression, one may take a small $\xi$ limit on both equations. This limit then allows us to integrate the above equations. We obtained the entanglement entropy in this limit as the following form :
\bea\label{analyticNex}
S^{\text{strip}}_{EE}&=&\frac{L}{4 G_N l}\biggl[\frac{2\,l}{\epsilon}+
\frac{\sqrt{2}\,\pi^2\,\Gamma(-\frac{1}{4})}{\Gamma(\frac{1}{4})^3}
+\frac{\pi\,r_h^2\,\gamma_1\,\Gamma(\frac{1}{4})}{\sqrt{2}\,\Gamma(-\frac{1}{4})^2\,\Gamma(\frac{7}{4})}l^2
+\frac{\pi\,r_h^3(2\,\gamma_1-4-\gamma_2)\Gamma(\frac{1}{4})}{\sqrt{2}\,\Gamma(-\frac{1}{4})^3}l^3
\nonumber
\\
&+&\frac{r_h^4(576\,\pi^3(3\,\gamma_1^2-4\,\gamma_2)\Gamma(\frac{3}{4})-5\sqrt{2}\,\gamma_1^2\,\Gamma(\frac{1}{4})^7)}{92160\,\pi^2\,\Gamma(\frac{3}{4})^5}l^4
\nonumber
\\
&+&\frac{8\,\pi^2\,r_h^5\,\gamma_1(2\,\gamma_1-4-\gamma_2)(\pi^3\,\Gamma(-\frac{1}{4})+ 48\,\Gamma(\frac{3}{4})^5)}{3\,\Gamma(-\frac{1}{4})^7\,\Gamma(\frac{3}{4})^6}l^5
+\mathcal{O}(l^6)
\biggr]~,
\eea
where
$r_h$ is given by $1/(l z_h)$. In addition, the $\gamma_1$ and $\gamma_2$ are defined by
\begin{eqnarray}
\gamma_1 \equiv \tilde{\beta}^2 z_h^2~~,~~\textcolor{black}{\gamma_2 \equiv \tilde{q}^2 z_h^4~~}.
\end{eqnarray}

\begin{figure}[t]
\centering
\begin{subfigure}[]
\centering
\includegraphics[width=0.35\textwidth]{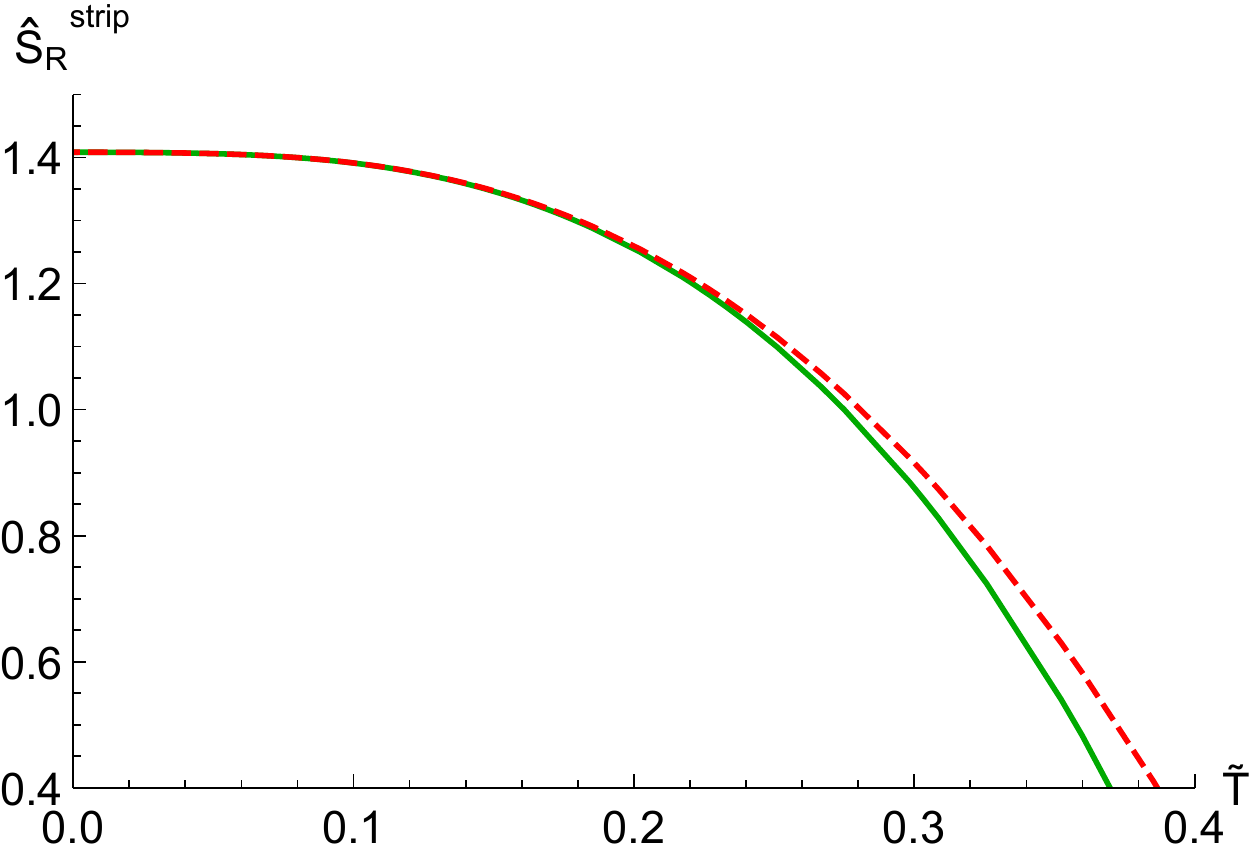}
\end{subfigure}\qquad
\begin{subfigure}[]
\centering
\includegraphics[width=0.35\textwidth]{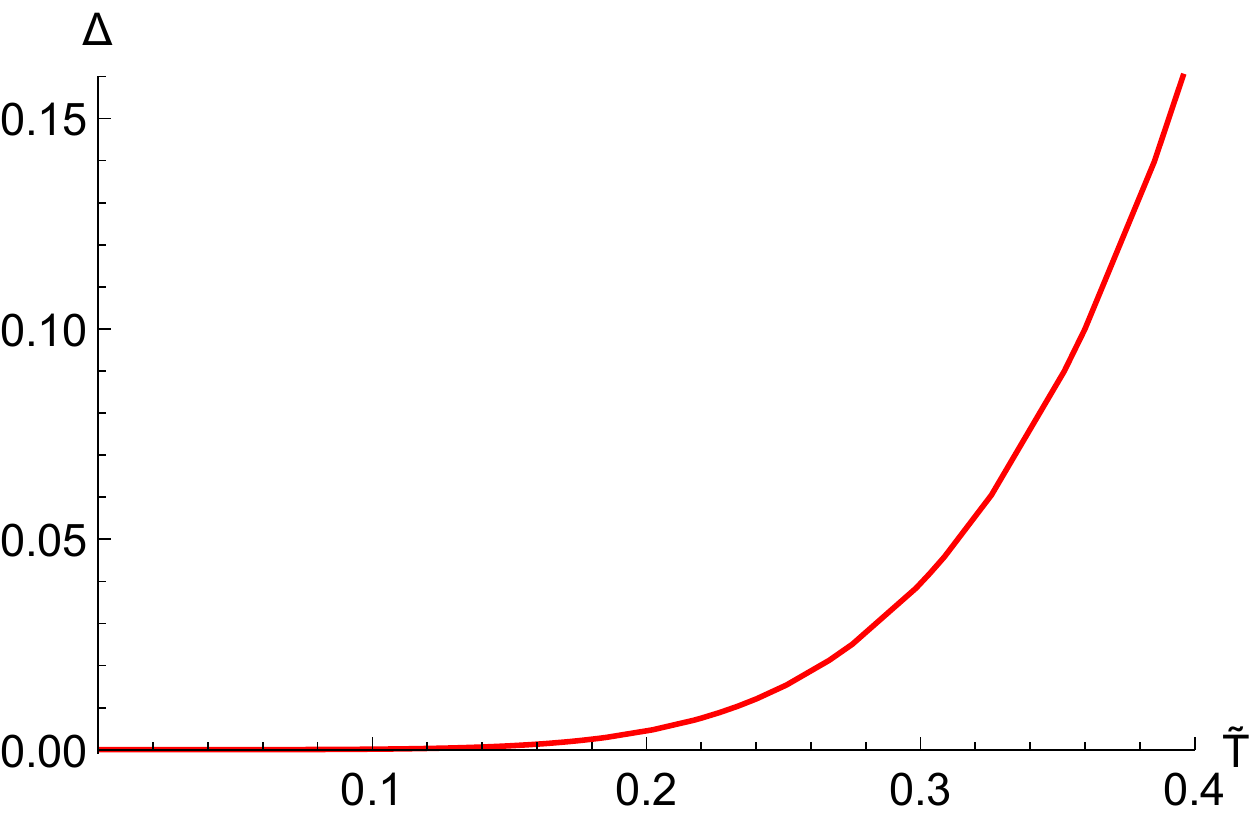}
\end{subfigure}
\begin{subfigure}[]
\centering
\includegraphics[width=0.35\textwidth]{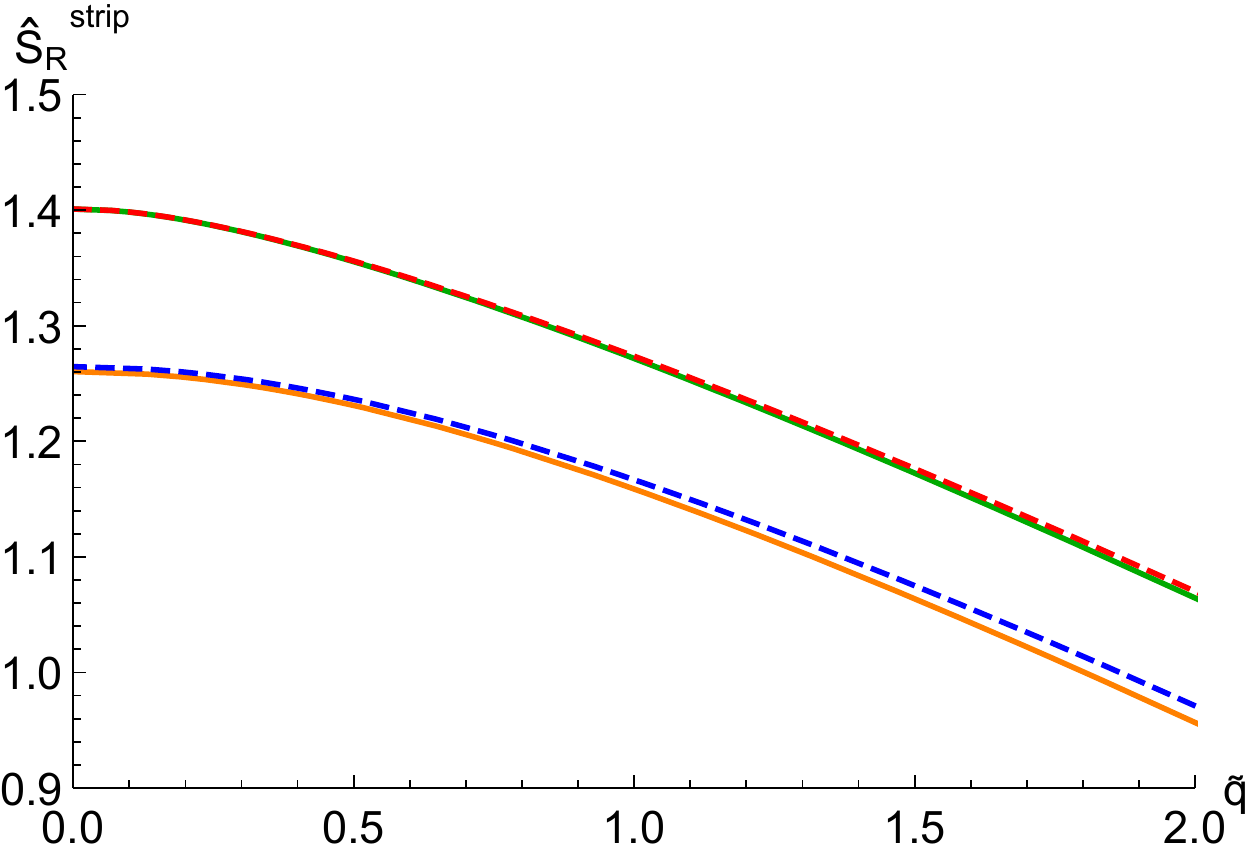}
\end{subfigure}\qquad
\begin{subfigure}[]
\centering
\includegraphics[width=0.35\textwidth]{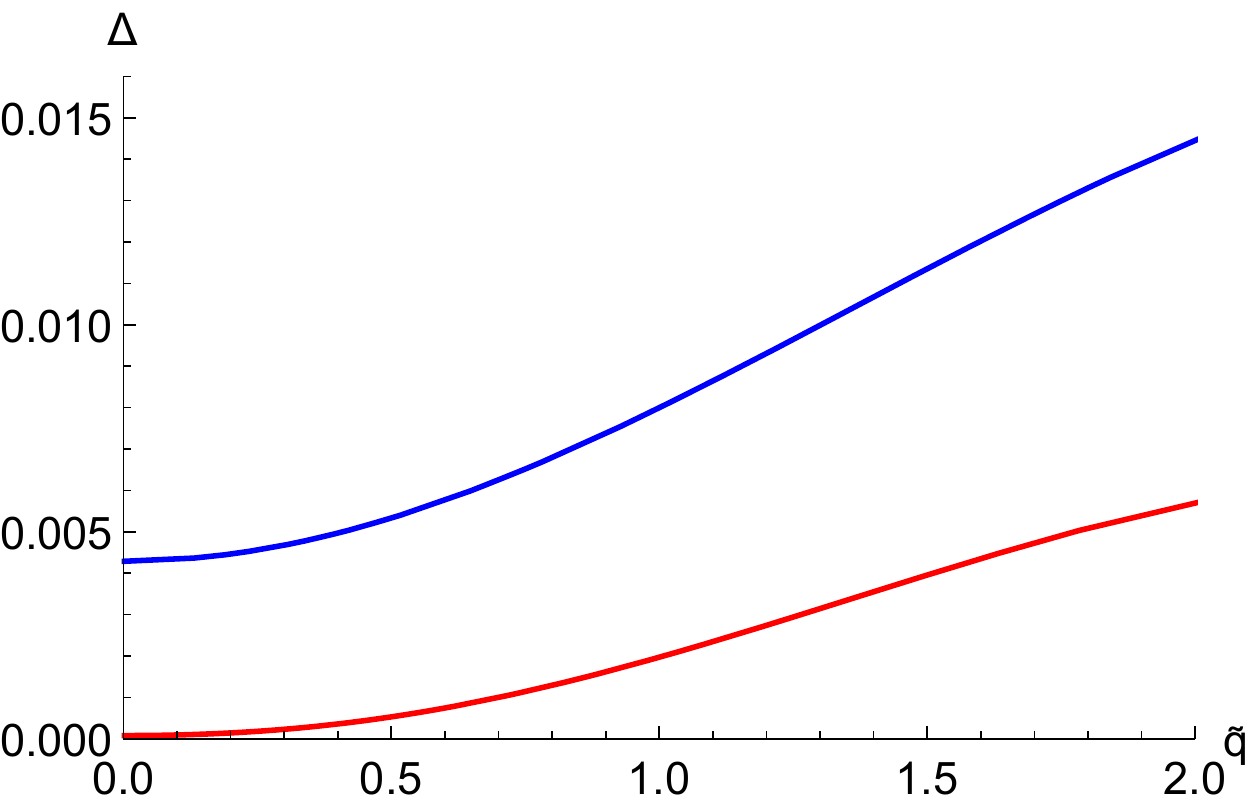}
\end{subfigure}
\caption{\textcolor{black}{Comparison of the numerical calculation(Dashed) and the analytic one(Solid) for $\tilde \beta = 0.2$: (a) is for a fixed charge density $\tilde q = 0.2$ and (b) describes the difference, $\Delta$, between two results. As $\tilde T$ grows, $\Delta$ also increases. (c) shows the case of a fixed temperature $\tilde T = 0.1$(Red-dashed, Green) and $\tilde T = 0.2$(Blue-dashed, Orange). In (d), $\Delta$ of the case $\tilde T = 0.2$(Blue) is larger than that of the case $\tilde T = 0.1$(Red). When $\tilde T$ has a small value, the analytic result gives a good approximation.}}\label{error}
\end{figure}

The physical meaning of this limit($\xi\ll 1$) is that the tip distance($z_*$) from the boundary is much smaller than that of the horizon of the black hole($z_h$). Roughly speaking, this limit is similar to $\frac{l}{z_h}\ll 1$, but it is not the equivalent limit in a strict sense. The analytic result (\ref{analyticNex}) is compared with the previous numerical calculations in Fig.\ref{error}.

The above result can also have an extremal limit. At zero temperature, the charge density is given by $\tilde q={\sqrt{12-2 z_h^2 \tilde{\beta}}}/{z_h^2}$. So the holographic entanglement entropy at zero temperature is determined to be:
\bea\label{analyticExt}
S^{\text{strip}}_{EE}\,=\frac{L}{4G_N l}&\biggl[&\frac{2\,l}{\epsilon}+
\frac{\sqrt{2}\,\pi^2\,\Gamma(-\frac{1}{4})}{\Gamma(\frac{1}{4})^3}
+\frac{64\,\pi^2\,\beta^2}{3\,\Gamma(-\frac{1}{4})^4}\,l^2
+\frac{16\,\pi^2\,r_h(4\,r_h^2-\beta^2)}{\Gamma(-\frac{1}{4})^4}\,l^3
\nonumber
\\
&&+\frac{8\,\pi\,(-160\,\pi^2\,\beta^4\,\Gamma(-\frac{3}{4})^2+(3\,\beta^4+8\,r_h^2\,\beta^2-48\,r_h^4)\Gamma(-\frac{1}{4})^6)}{5\,\Gamma(-\frac{1}{4})^{10}}\,l^4
\nonumber
\\
&&-\frac{ 8\pi^2\,r_h\,\beta^2(4r_h^2-\beta^2)(\pi^3\,\Gamma(-\frac{1}{4})+48\,\Gamma(\frac{3}{4})^5)}{\Gamma(-\frac{1}{4})^7\,\Gamma(\frac{3}{4})^5\,\Gamma(\frac{7}{4})}l^5
+\mathcal{O}(l^6)
~~~\biggr]~.
\eea
This shows the effect of momentum relaxation on the holographic entanglement entropy when the temperature is zero.

The above two results quantitatively indicate contribution of impurity to the entanglement entropy. It is very interesting to compare these results with the effects of impurities in weakly coupled field theories. On the other hand, the leading contribution in (\ref{analyticExt}) is proportional to $\beta^2$. It is remarkable that impurities {\color{black}reduce} entanglement entropy in a very narrow strip and this result can be compared to known results in condensed matter physics, our system is different from systems studied in condensed matter physics though. See \cite{Bidzhiev:2017eye} and its references for recent articles. One may read \cite{Laflorencie:2015eck} for a recent review.

{\color{black}
As a final remark, we may think the strip case with a rotated electric field. This is achieved by replacing the subindex $x$ in  (\ref{def:fluctuationsfields}) by the index $i$ denoting $(x,y)$ with nonvanishing $y$-components. However, the resulting action is still the same as the previous one up to the redefinition of the coordinate in \ref{mapping}. This fact implies that the rotation of the electric field doesn't give any nontrivial contribution to the action of the minimal surface. In section \ref{Section:wedge}, we will consider the rotated electric field with a wedge entangling region which, even at the linear order, leads to a nontrivial effect on the entanglement entropy.
}

\subsection{ A Comparable Deformation of ABJM theory  }

{\color{black}Our results (\ref{analyticNex}) and (\ref{analyticExt}) tell us how the entanglement entropy changes under deforming the CFT vacuum with operators dual to the axion field and the bulk U(1) gauge field in (\ref{ActionGravity}). The background solution (\ref{Ansatz00}) describes the deformation more explicitly. Since we do not turn on any magnetic field, the operator dual to the gauge field is nothing but the charge density operator $\mathcal{Q}$ and the source is given by the chemical potential. {\color{black} In addition, the axion field which consists of two massless real scalar fields are dual to dimension 3 operators $\mathcal{O}^{\Delta=3}_I$.} 

One of well-known examples of gauge/gravity correspondence is the duality between ABJM theory and $AdS_4\times S^7/\mathbb{Z}_k$ \cite{Aharony:2008ug}. If we take the planar limit of ABJM theory, then the dual geometry becomes $AdS_4\times \mathbb{CP}_3$. Therefore the ABJM theory becomes dual to a 4-dimensional supergravity model that is obtained from the type IIA supergravity compactified on $\mathbb{CP}_3$. Thus one can find candidates for the fields in (\ref{ActionGravity}). The bulk local gauge symmetry appears as a global symmetry in the boundary. The ABJM theory has $SU(4)$ R-symmetry and the $U(1)$ symmetry can be embedded in the R-symmetry. So it is possible to identify $\mathcal{Q}$ with the charge operator of the diagonal $U(1)$ of $SU(4)$ R-symmetry, $\mathcal{Q}^R$. In addition there are many chiral primary scalar operators with dimension 3, which are candidates for $\mathcal{O}^{\Delta=3}_I$. See \cite{Bak:2010yd} and table 1 in \cite{Bak:2010ry}. Therefore the following deformation of the ABJM theory is dual to (\ref{ActionGravity}):
\begin{align}
S' = S_{ABJM} + \int d^3 x \left( \mu \mathcal{Q}^R + \sum_{I=1}^2   \mathcal{S}^I_{\Delta=3}\,\mathcal{O}^{\Delta=3}_I \right)~,
\end{align}      
where $S_{ABJM}$ is the action of the ABJM theory and $\mathcal{S}^I_{\Delta=3}$ are source for the scalar operators in the ABJM theory. And it is given by $\mathcal{S}^I_{\Delta=3} = \left( \beta\, x ,\beta\, y \right)$. There could be some difference coming from nonlinear effects of compactification on $\mathbb{CP}_3$. If, however, we consider small deformations and small size of the entangling region, the correction can be negligible. Then, it is valuable to compare the entanglement entropy with the above deformed action to (\ref{analyticNex}) and (\ref{analyticExt}). We hope to see a comparison of our result and the entanglement entropy obtained by a developing technique of supersymmetric field theory.  }

\section{Holographic Entanglement Entropy in Electric Field: Wedge Entangling Region}\label{Section:wedge}

In this section, we consider the holographic entanglement entropy with wedge entangling regions. This type of entangling surfaces has been studied in \cite{Bueno:2015rda,Bueno:2015xda,Faulkner:2015csl,Miao:2015dua} to consider the corner contribution of entanglement entropy. It turns out that this corner contribution contains universal information of conformal field theory, which will be discussed later.

\begin{figure}
\centering
(a)\includegraphics[width=0.4\textwidth]{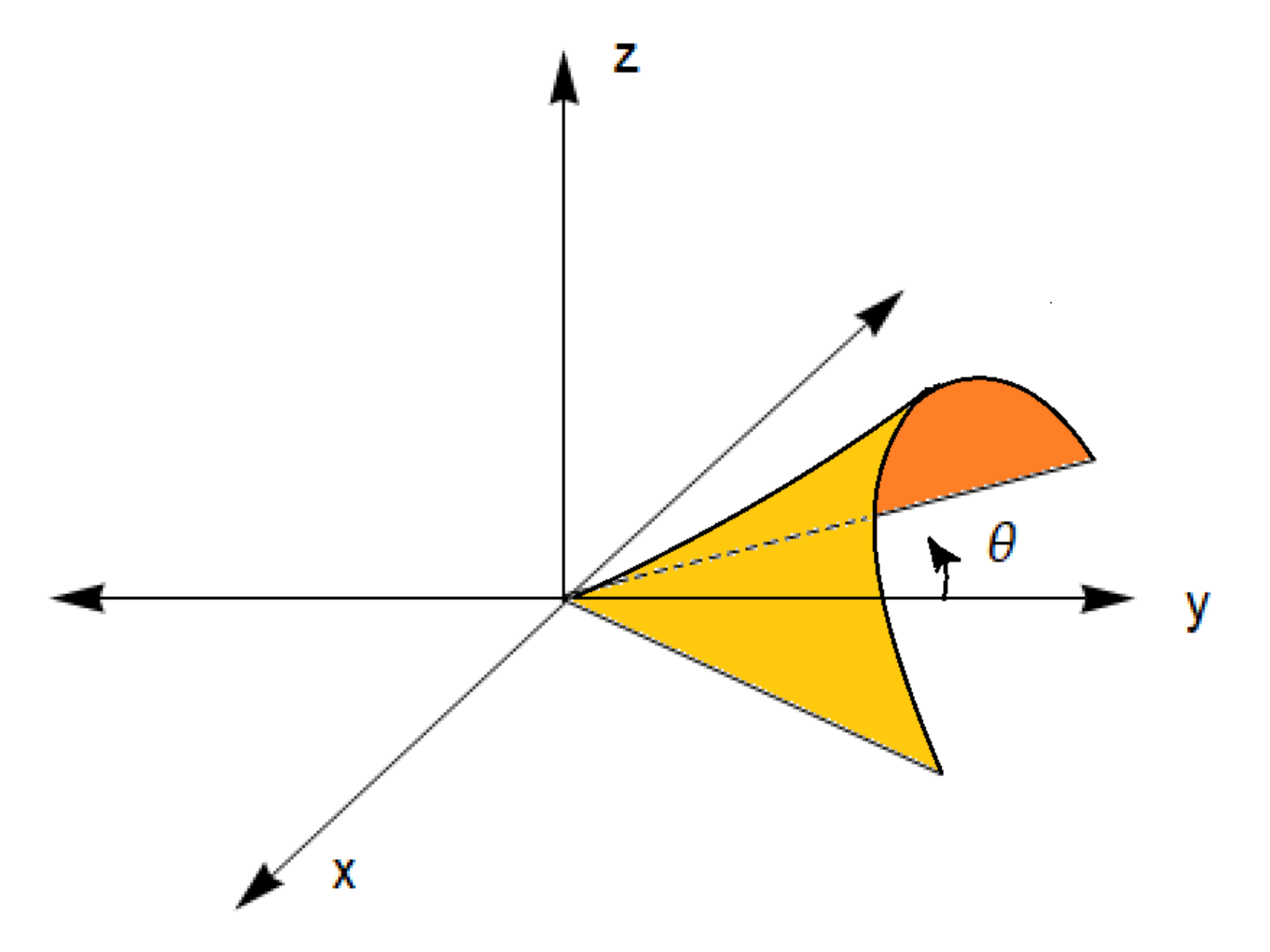}
(b)\includegraphics[width=0.4\textwidth]{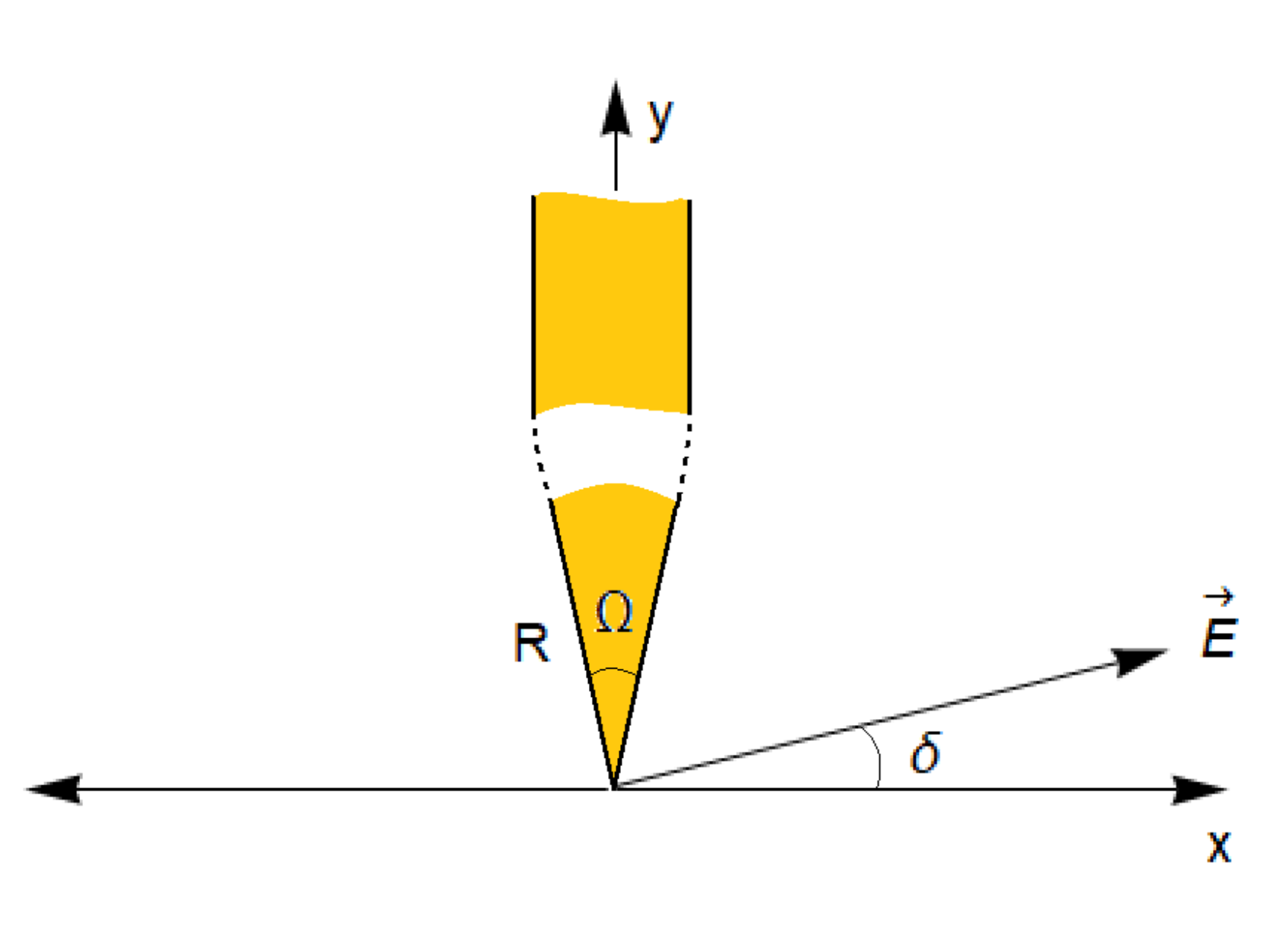}
\caption{Cartoons for a minimal surface(a) and its top view(b) : We regard the wedge type minimal surface as a part of a larger structure, such as a tail of a strip type minimal surface. We have depicted that imagination in the top view(b).}\label{fig2}
\end{figure}

See Fig.\,\ref{fig2} for a cartoon of a minimal surface anchored to a wedge region. It is useful to take polar coordinates $(\rho,\theta)$ in the $x-y$ plane as the coordinates of the minimal surface. Our main concern is how much the holographic entanglement entropy changes at the linear level under the weak electric field. To see this linear response, the direction of the electric field must be rotated by an angle $\delta$, and one of the Cartesian coordinates $x$ is given by $x = - \rho \sin (\theta-\delta)$. The need for the rotation was inferred by lessons of the previous section.

Considering the above configuration, the background geometry can be  written as follows:
\begin{eqnarray}\label{metric2}
ds^2&=&-U(r)dt^2+\frac{dr^2}{U(r)}+r^2(d\rho^2+\rho^2 d\theta^2)\\&&-2\lambda\biggl(
g_{tx}(r)dt\,+g_{rx}(r)dr
\biggr)\left( \rho  \cos (\theta -\delta )d \theta + \sin (\theta -\delta )d\rho\right)+ \mathcal{O}\left(\lambda^2\right).\nonumber
\end{eqnarray}
And we may identify the coordinates of the minimal surface as the following way:
\begin{eqnarray}
t=0~,~r=r(\sigma_1,\sigma_2)=1/{\tilde{z}}(\sigma_1,\sigma_2)~,~\rho=\sigma_1~.~\theta=\sigma_2
\end{eqnarray}
Then the action for the minimal surface is given by
\begin{eqnarray}
\mathcal{A} = \int_{-\Omega/2}^{\Omega/2} d\theta \int_0^\infty d\rho \left(\frac{\rho}{{\tilde{z}}^2}  \sqrt{1+\frac{{\tilde{z}}'^2+\rho ^2 \dot{{\tilde{z}}}^2}{\rho ^2 {\tilde{z}}^2 f({\tilde{z}})}}  + \lambda\, \frac{\mathbb{H}_{rx}\left({\tilde{z}}' \cos (\delta -\theta )-\rho  \dot{{\tilde{z}}} \sin (\delta -\theta )\right) }{{\tilde{z}}^2 f({\tilde{z}}) \sqrt{1+\frac{{\tilde{z}}'^2+\rho ^2 \dot{{\tilde{z}}}^2}{\rho ^2 {\tilde{z}}^2 f({\tilde{z}})}}}   \right)~~.
\end{eqnarray}
We are interested only in a variation of the action under the electric field and regard the wedge as a part of a larger minimal surface like the cartoon (b) in Fig. \ref{fig2}. So we consider a part of the minimal surface with the region, $\rho < R$ and we will see how the surface changes for this finite region. By this reason, the integration for the radial coordinate $\rho$ is replaced by the integration between 0 to $R$. Furthermore, we will take a convenient choice of parameters obtained by a scaling with respect to $\Omega$ :
\begin{eqnarray}
\sigma = \theta/\Omega~~,~~{\tilde{z}} = \Omega w ~~,~~\tilde \beta = \Omega \beta~~,~~\tilde M = \Omega^3 M ~~\tilde q = \Omega^2 q~~.
\end{eqnarray}
Then, the action becomes
\begin{eqnarray}
\mathcal{A} = \frac{1}{\Omega}\int_{-1/2}^{1/2}d\sigma \int_0^R  & d\rho &\left( \frac{\rho}{w^2}\sqrt{1+\frac{w'^2+{\Omega^2}\rho ^2 \dot{w}^2}{\rho ^2 w^2 f(w)}}\right.\nonumber\\
&&~~~~+\left.\lambda \Omega^2\frac{\mathbb{H}_{rx}\left(w' \cos (\delta -\Omega\,\sigma )-{\Omega}\,\rho\,\dot{w} \sin (\delta -\Omega\,\sigma )\right) }{w^2 f(w) \sqrt{1+\frac{w'^2+{\Omega^2}\rho ^2 \dot{w}^2}{\rho ^2 w^2 f(w)}}} \right),
\end{eqnarray}
where $f(w)$ is given by
\begin{eqnarray}
f(w) = \frac{1}{w^2}-\frac{\tilde \beta^2}{2}-\tilde M w  +  \frac{\tilde q^2}{4} w^2~.
\end{eqnarray}
The above action determines the minimal surface by solving the equation of motion for $w(\rho,\sigma)$. In general, the equation is not easy to solve.


Now let us take a particular limit considering a sharp wedge. Then one can use $\Omega$ as another small parameter. The action up to the quadratic order in $\Omega$ is given by
\begin{eqnarray}
\mathcal{A}=\frac{1}{\Omega}\int_{-1/2}^{1/2}d\sigma \int_0^R d\rho \left(  \sum_{n=0}^{3} \mathcal{L}^{0th}_{(n)} \Omega^n + \lambda\, \sum_{n=2}^{3}\mathcal{L}_{(n)}^{1st}\Omega^n + \mathcal{O} \left( \Omega^4\right)   \right)~,
\end{eqnarray}
where
\begin{eqnarray}
\mathcal{L}^{0th}_{(0)}&=&\frac{\rho  \sqrt{\frac{w'^2}{\rho ^2}+1}}{w^2},~~\mathcal{L}^{0th}_{(1)}=0,~~\mathcal{L}^{0th}_{(2)}=\frac{\left(\beta ^2 w^2 \left(w'\right)^2+2 \rho ^2 \dot{w}^2\right)}{4 \rho  w^2 \sqrt{\frac{\left(w'\right)^2}{\rho ^2}+1}},~~
\mathcal{L}^{0th}_{(3)}=\frac{M w  w'^2}{2 \rho  \sqrt{\frac{w'^2}{\rho ^2}+1}}~,
\end{eqnarray}
\begin{eqnarray}
\mathcal{L}^{1st}_{(2)} =\mathbb{H}_{rx} \frac{w' \cos\delta }{\sqrt{\frac{w'^2}{\rho ^2}+1}}~~~\text{and}~~~\mathcal{L}^{1st}_{(3)}=\mathbb{H}_{rx}\frac{   \left(\sigma  w'-\rho \dot{w}\right)\sin\delta}{\sqrt{\frac{w'^2}{\rho ^2}+1}}~.
\end{eqnarray}
Since $\Omega$ is considered small, the solution $w$ is closed to the form of $\rho$ times a function of $\sigma$. Thus we assume that $w$ can be represented by a polynomial of $\rho$. Under this assumption, we found a solution of the equation of motion as the following form:
\begin{eqnarray}
w(\rho,\sigma) &=& w_0(\rho,\sigma) + \lambda\, w_1(\rho,\sigma) \nonumber\\ &=&\rho\,  h_{0,1}(\sigma ) +\Omega^2 \left(\sum _{q=1}^3  h_{2,q}(\sigma )\rho^q\right)+\Omega^3  h_{3,4}(\sigma )\rho^4 \nonumber
%
%
\\&&+\lambda\,\mathbb{H}_{rx} \left(  \Omega^2 g_{2,3}(\sigma )\rho^3\cos\delta+ \Omega^3 g_{3,3}(\sigma )\rho^3 \sin\delta\right)~.\label{solw}
\end{eqnarray}
The differential equations for the functions, $h_{p,q}$ and $g_{p,q}$ are given in Appendix A. In addition, the regularized area is given by
\begin{eqnarray}
\mathcal{A}_{reg} = \frac{1}{\Omega}\int_{\sigma_{min}}^{\sigma_{max}} d\sigma \int_{\alpha(\sigma)+ \lambda\,\beta(\sigma)}^R d\rho \left(  \sum_{n=0}^{3} \mathcal{L}^{0th}_{(n)} \Omega^n + \lambda\, \sum_{n=2}^{3}\mathcal{L}_{(n)}^{1st}\Omega^n \right)~,
\end{eqnarray}
where $\alpha(\sigma)+ \lambda\,\beta(\sigma)$ is the cut-off line obtained from ${\tilde{z}}=\epsilon$ and the explicit form is given in Appendix A.
In this regard, $\sigma_{max}$ and $\sigma_{min}$ denote the maximum and minimum values of $\sigma$ satisfying $\alpha(\sigma)+ \lambda\,\beta(\sigma)=R$. And they can be expanded as $
\sigma_{max,min}\sim\sigma_{max,min}^{(0)}+\lambda\,\sigma_{max,min}^{(1)}$, respectively. The regularised action to the linear order in $\lambda$ is
\begin{eqnarray}
\mathcal{A}_{reg} &=& \frac{1}{\Omega}\int_{\sigma_{min}^{(0)}}^{\sigma_{max}^{(0)}} d\sigma \int_{\alpha(\sigma)}^R d\rho\,  \left[  \left(  \sum_{n=0}^{3} \mathcal{L}^{0th}_{(n)} \Omega^n \right)_{w_0 + \lambda w_1}
+ \lambda \left( \sum_{n=2}^{3}\mathcal{L}_{(n)}^{1st}\Omega^n\right)_{w_0} \right]  \nonumber\\
&&+\, \frac{1}{\Omega} \left(\int_{\sigma_{min}}^{\sigma_{max}} d\sigma \int_{\alpha(\sigma)+ \lambda\beta(\sigma)}^{R} d\rho - \int_{\sigma^{(0)}_{min}}^{\sigma^{(0)}_{max}} d\sigma \int_{\alpha(\sigma)}^{R} d\rho \right) \left(  \sum_{n=0}^{3} \mathcal{L}^{0th}_{(n)} \Omega^n  \right)_{w_0} +\mathcal{O}\left(\lambda^2\right) ~.
\end{eqnarray}

Plugging (\ref{solw}) into the above action, we can find the on-shell action as follows:
\begin{eqnarray}
\mathcal{A}_{reg}&\sim & \frac{1}{\Omega}\int_{\sigma_{min}^{(0)}}^{\sigma_{max}^{(0)}} d\sigma \int_{\alpha(\sigma)}^R d\rho\,  \left(  \sum_{n=0}^{3} \mathcal{L}^{0th}_{(n)} \Omega^n \right)_{w_0} \nonumber  \\
&&-\frac{\lambda\, \mathbb{H}_{rx}}{\Omega}\cos\delta\,\int_{-\sigma_{max}^{(0)}}^{\sigma_{max}^{(0)}} d\sigma \int_{\alpha(\sigma)}^R d\rho\,\frac{\rho\, \Omega^2 \left(2 g_{2,3} \left(h_{0,1}'^2+1\right)-h_{0,1} g_{2,3}' h_{0,1}'\right)}{h_{0,1}^3 \sqrt{h_{0,1}'^2+1}}\nonumber\\
&&-\frac{\lambda\, \mathbb{H}_{rx}}{\Omega}\sin\delta\,\int_{-\sigma_{max}^{(0)}}^{\sigma_{max}^{(0)}} d\sigma \int_{\alpha(\sigma)}^R d\rho\,\frac{\rho\, \Omega^3 \left(2 g_{3,3} \left(h_{0,1}'^2+1\right)-h_{0,1} g_{3,3}' h_{0,1}'\right)}{h_{0,1}^3 \sqrt{h_{0,1}'^2+1}}\nonumber\\
&&+\frac{\lambda\, \mathbb{H}_{rx}}{\Omega}\int_{-\sigma_{max}^{(0)}}^{\sigma_{max}^{(0)}} d\sigma \int_{\alpha(\sigma)}^R d\rho\,\rho\left(\frac{\Omega^2 h_{0,1}'\cos\delta}{\sqrt{h_{0,1}'^2+1}}+ \frac{\Omega^3 \left(\sigma  h_{0,1}'-h_{0,1}\right) \sin\delta}{\sqrt{h_{0,1}'^2+1}}\right) \nonumber\\
&&+\frac{\lambda}{\Omega}\left( \sigma^{(1)}_{max}-\sigma^{(1)}_{min} \right)\left(\int_{\alpha(\sigma)}^{R}d\rho\left(  \sum_{n=0}^{3} \mathcal{L}^{0th}_{(n)} \Omega^n \right)_{w_0} \right)_{\sigma = \sigma^{(0)}_{max}}
\nonumber\\
&&-\frac{\lambda}{\Omega} \int_{-\sigma_{max}^{(0)}}^{\sigma_{max}^{(0)}}d\sigma \beta(\sigma) \left(  \sum_{n=0}^{3} \mathcal{L}^{0th}_{(n)} \Omega^n \right)_{\rho=\alpha(\sigma)}  ~~,\label{AregW}
\end{eqnarray}
where we used $\sigma_{min}^{(0)}=-\sigma_{max}^{(0)}$. And the first order correction for $\sigma_{min}$ and $\sigma_{max}$ is given by
\begin{eqnarray}
\sigma_{min,max}^{(1)} = - \frac{\beta(\sigma_{min,max}^{(0)})}{\alpha'(\sigma_{min,max}^{(0)})}~~.
\end{eqnarray}
This can be easily derived from $R=\alpha(\sigma_{min,max})+ \lambda\,\beta(\sigma_{min,max})$\,.

{\color{black}The zeroth order part in $\lambda$ is given by the first line in (\ref{AregW}) . 
In fact this part shows universal features of underlying CFTs \cite{Bueno:2015rda,Bueno:2015xda,Bueno:2015qya,Sierens:2017eun}, which are summarized as follows. This term has two kinds of divergent terms. One is the usual UV divergent term which is proportional to $1/\epsilon$ that appears commonly in 2+1 dimensional field theory systems. One can see the term also in the strip case, (\ref{analyticNex}). In addition to this, another log-divergent term shows up. Since this term is coming from the singular corner of the wedge, it is called the corner contribution.  The general structure of the holographic entanglement entropy is as follows:
\begin{align}
S^{(0)}_{EE}=\frac{1}{4 G_N}\mathcal{A}^{(0)}_{reg}  = \mathbb{B}\,\frac{\mathbb{L}}{\epsilon} - \mathbb{A}(\Omega)\, \log \frac{\mathbb{L}}{\epsilon} + \mathbb{C}^{(0)} + \mathcal{O}\left( \frac{\epsilon}{\mathbb{L}} \right)~~,
\end{align} 
where $\mathbb{L}$ is a typical length scale characterized by the size of the entangling region. Even though the constant $\mathbb{B}$ governs the leading term, it crucially depends on UV regulators. $\mathbb{C}^{(0)}$ is the finite part of entanglement entropy. The most interesting quantity is the coefficient of log-term, $\mathbb{A}(\Omega)$. The characteristic features of $\mathbb{A}$ are determined by two limits of $\Omega$:
\begin{align}
\lim_{\Omega\to 0} \mathbb{A} \sim \frac{\kappa_1}{\Omega}~~,~~\lim_{\Omega\to \pi} \mathbb{A} \sim \sigma_1 (\pi-\Omega)^2~~,
\end{align}
where $\kappa_1$ is conjectured to be related to, so called, entropic c-function and $\sigma_1$ is given by $\sigma_1 = \frac{\pi^2}{24}C_T$. $C_T$ is the central charge appearing in the vacuum two point function as follows:
\begin{align}
\left< T_{\mu\nu}(x)\, T_{\alpha\beta}(0) \right> = \frac{C_T}{|x|^{4}}\, \mathcal{I}_{\mu\nu,\alpha\beta}(x)~~,
\end{align}
where $\mathcal{I}_{\mu\nu,\alpha\beta}$ is a dimesionless tensor which is completely fixed by conformal symmetry \cite{Osborn:1993cr}. It is worth noting that the coefficient of the log-divergent term $\mathbb{A}$ is UV regulator independent, as opposed to $\mathbb{B}$. In addition these results were generalized to the Reyni entropy. See \cite{Bueno:2015xda} and references therein.  
 }

{\color{black}Even though the above zeroth order part is very interesting}, our main interest here is the linear order variation. {\color{black} So we focus on the first order part.} If we write down only the first order part using some algebra in Appendix B, the linear action in $\lambda$ is :
\begin{eqnarray}
\delta \mathcal{A}_{reg}= -\lambda\, \mathbb{H}_{rx}\Omega^2 \,\sin\delta\, \int_{-\sigma^{(0)}_{max}}^{\sigma^{(0)}_{max}}d\sigma\left( \mathcal{I}_1 + \mathcal{I}_2 \right)+\mathcal O \left( \Omega^3 \right)~~,
\end{eqnarray}
where
\begin{eqnarray}
%
%
\mathcal{I}_1 &=&\frac{\left(R^2-\alpha (\sigma )^2\right) \left(2 g_{3,3}(\sigma ) \left(h_{0,1}'(\sigma ){}^2+1\right)-h_{0,1}(\sigma ) g_{3,3}'(\sigma ) h_{0,1}'(\sigma )\right)}{2 h_{0,1}(\sigma ){}^3 \sqrt{h_{0,1}'(\sigma ){}^2+1}}~,\\
%
\mathcal{I}_2 &=&  \left(R^2-\alpha (\sigma)^2\right) \left( - \frac{\lambda  \left(\sigma  h_{0,1}'(\sigma )-h_{0,1}(\sigma )\right)}{2 \sqrt{h_{0,1}'(\sigma )^2+1}} \right)~.
\end{eqnarray}
Here the contribution of $\alpha(\sigma)^2$ term to the integration
is much smaller than the contribution of $R^2$ term. See Appendix B for the explicit form of $\alpha(\sigma)$ that is order of $\mathcal{E}$. Thus we drop the $\alpha(\sigma)^2$ term.

If we drop out the discussed higher orders in $\epsilon$ and the integrations of some odd functions\footnote{One can easily see that all $h_{p,q}$'s and $g_{3,3}$ are even functions and $g_{2,3}$ is an odd function from the equations in Appendix A.}, the leading change of the regularized minimal surface to the external electric field is given by
\begin{eqnarray}
&&\delta \mathcal{A}_{reg}\label{dA}\\\nonumber
&&~~=\lambda\, R^2\, \Omega^2\,\mathbb{H}_{rx}\,\sin\delta \int_{-\sigma^{(0)}_{max}}^{\sigma_{max}^{(0)}}d\sigma\frac{-\sigma  h(\sigma)^3 h'(\sigma )+h(\sigma)^4+h(\sigma ) h'(\sigma ) g'(\sigma )-2 g(\sigma ) \left(h'(\sigma )^2+1\right)}{2 h(\sigma )^3 \sqrt{h'(\sigma )^2+1}}~~,
\end{eqnarray}
where $h(\sigma)=h_{0,1}(\sigma)$ and $g(\sigma)=- g_{3,3}(\sigma)$. In addition $g$ and $h$ satisfy the following equations :
\begin{eqnarray}
&&h'' =-\frac{2 \left(h'^2+1\right)}{h }~,\\
&&g'' =-\frac{4 g'  h' }{h }+\frac{2g\left(1+h'^2\right)}{h^2} -2 h^2+6 \sigma  h h' +4  h^2 h'^2~.
\end{eqnarray}
One can solve the above equations numerically and perform the numerical integration in (\ref{dA}). Finally, we can get the linear response of the holographic entanglement entropy to the electric field as follows.
\begin{eqnarray}
\frac{\delta \mathcal{S}_{EE}}{\delta E_x} = \frac{R^2\, \Omega^2\,}{4 G_N}\mathcal{N} \frac{\bar\alpha}{4\pi}\sin\delta~~,
\end{eqnarray}
where $\mathcal N$ is the integration in (\ref{dA}). The integrand of $\mathcal{N}$ is a positive function over $-\frac{1}{2} \leq \sigma \leq \frac{1}{2}$, thus $\mathcal{N}$ does not vanish and it is about $0.05$. Therefore, the linear response of the holographic entanglement entropy to the electric field is proportional to the thermoelectric conductivity $\bar\alpha$ which is a measurable quantity.

This is the first result for the relation between the entanglement {\color{black} response} and transport in the gauge/gravity duality. In condensed matter community, related physics is a quite active research topic and has produced many interesting results. See \cite{Laflorencie:2015eck} for a review. The origin of the linearity of the thermal conductivity $\bar \alpha$ comes from turning on the metric component $g_{rx}$ by the external electric field. Together with a reasonable gauge choice, this component describes the thermoelectric current. This hallmark is universal in {\color{black} various holographic models with momentum relaxation. Therefore, the above linear relation between the entanglement entropy variation and the thermoelectric coefficient is quite noticeable as a universal feature of entanglement in strongly coupled systems. Naturally, one can ask the reason why the leading effect is from the heat current mediated by the thermoelectric coefficient not from the electric current by the electric conductivity. This question is very difficult to answer because such an effect is closely related to the Fermi surface. We leave this investigation for our future work.}

{\color{black} 
The measurement of entanglement is a very intriguing topic in various areas of condensed matter physics. The first measurement of the second R{\'e}nyi entropy was accomplished in \cite{islam}. This work was inspired by
\cite{Hrodecki:2002,Alves:2004}. Except for this measurement, there are various trials to detect entanglement measure. Although the accomplished detections were seen through few-body correlations, there are many discussions and proposals on extension of the experiments. Among them, a suitable proposal to our result is based on quantum antiferromagnets. See 
\cite{Song:2012}. In the proposal, the entanglement measure can be encoded in magnetization of a system. A circular superconducting plate covers on both sides of an antiferromagnet sheet. By the Meissner effect, the plate plays a role of dividing entangling regions. Assuming that such an experiment is realized, we propose an experiment with a wedge type superconducting plate and applying an rotated electric field along a tangential direction. Then, one may check whether resulting entanglement entropy data is related to the thermoelectric coefficient $\bar\alpha$ or not. Our idea is a bit rough, but worth thinking about more.

}

\section{Discussion}

In this work, we have studied the entanglement entropy affected by the external electric field. Though the entanglement entropy has usually been considered as an important and fundamental quantity representing a variety of quantum natures, it still remains a big issue to measure the quantum entanglement entropy in the laboratory. One of the main goals of this work is how we can relate the quantum entanglement entropy to the measurable quantities like the transport coefficients. This would be useful to figure out the relation between the quantum phenomena and various macroscopic quantities. Moreover, the present work may provide new intuitions about how to measure the entanglement entropy in the laboratory.

More specifically, we have taken into account the entanglement entropy of the strip and wedge entangling region with turning on the electric field and momentum relaxation. If there is no momentum relaxation which breaks the translation symmetry, the DC conductivity usually has an infinite value.
In order to obtain a finite conductivity in the zero frequency limit, we first considered the geometry with the non-vanishing momentum relaxation which resembles introducing impurities to the dual field theory. This momentum relaxation usually modifies the electric and thermoelectric conductivities. Theoretically, those changes of the transport coefficients can be governed by the response theory and we can easily measure such changes experimentally in the laboratory. In this situation, we can ask how the entanglement entropy is affected by the external electric field and what is the relation between the modified entanglement entropy and transport coefficients. In this paper, we showed how the entanglement entropy modified by the external electric field using the holographic method. It turns out that the entanglement entropy change can be connected to the transport coefficient, especially the thermoelectric conductivity.

For the strip case, we found that turning on the external electric field tilted the minimal surface as expected. In the holographic model, the area of the minimal surface is directly associated with the entanglement entropy. Thus, we may expect the change of the entanglement entropy due to the external electric field. We found that such a change of the entanglement entropy does not occur at least in the linear response theory. This is because the holographic formula of the entanglement entropy in (\ref{eq:HEEperpendicular0}) has an invariant form under the parity transformation like $x \to -x$. Despite the tilted minimal surface, the invariance under the parity transformation does not allow the change of the entanglement entropy at least at the linear order. However, the higher order corrections can affect the resulting entanglement entropy. In fact, we explicitly showed that the second order correction caused by the tilted minimal surface can change the entanglement entropy, although we did not regard an additional contribution caused by the background metric deformation. However, we did not present the result in this note because we leave it as part of future works. The additional contribution is related to the response theory at the second order and we hope to report more results on this issue in future works. Anyway, the results in the strip entangling region showed the fact that the change of the entanglement entropy can be represented in terms of transport coefficients.

In order to get more clear and explicit relation at the linear order, we further considered the entanglement entropy in the wedge entangling region which is utilized to extract universal information about the corner contribution. As shown in the strip case, the entanglement entropy invariant under the parity transformation does not give any nontrivial contribution at the linear order, so that we took into account the external electric field which is rotated with an arbitrary angle. Since the external electric field with an arbitrary rotation usually breaks the parity invariance, one can expect the nontrivial contribution to the entanglement entropy even at the linear order. We showed with explicit calculations that the linear order correction to entanglement entropy really occurs in this setup. Intriguingly, we furthermore found that the change of the entanglement entropy is directly related to the thermoelectric conductivity. This is the first example showing how the entanglement entropy can be connected to the transport coefficients. {\color{black}In addition we provided an rough experimental idea for our result of the wedge case and we also described a comparable deformation in the ABJM theory to the strip case.} In future works, we hope to report more evidences and understanding of the underlying structure of the connection between the entanglement entropy and transport coefficients.

\vspace{0.5cm}

{\bf Acknowledgement}

K. Kim appreciates APCTP for its hospitality during completion of this work. This work was supported by the Korea Ministry of Education, Science and Technology, Gyeongsangbuk-Do and Pohang City. C. Park was also supported by Basic Science Research Program through the National Research Foundation of Korea funded by the Ministry of Education (NRF-2016R1D1A1B03932371). J. Lee was also supported by the same fund with grant number NRF-2016R1A6A3A01010320. K.Kim and B.Ahn were supported by the same fund with grant number NRF-2015R1D1A1A01058220. {\color{black}This research was partially(B. Ahn) supported by the National Research Foundation of Korea(NRF) grant with the grant number NRF-2016R1D1A1A09917598, by the Yonsei University Future-leading Research Initiative of 2017(2017-22-0098) and by the Graduate School of YONSEI University Research Scholarship Grants in 2000.}
\vspace{0cm}

{\center \section*{Appendix}}

\section*{A. Equations for $h_{p,q}$ and $g_{p,q}$}

\begin{eqnarray}
h_{0,1}''&=&-\frac{2 \left(\left(h_{0,1}'\right){}^2+1\right)}{h_{0,1}}\\
h_{2,1}''&=&\frac{h_{0,1}^3 \left(2 \left(h_{0,1}'\right){}^2-1\right)-4 h_{0,1} h_{0,1}' h_{2,1}'+2 h_{2,1} \left(\left(h_{0,1}'\right){}^2+1\right)}{h_{0,1}^2}\\
h_{2,2}''&=&\frac{2 \left(h_{2,2} \left(\left(h_{0,1}'\right){}^2+1\right)-2 h_{0,1} h_{0,1}' h_{2,2}'\right)}{h_{0,1}^2}\\
h_{2,3}''&=&-\frac{\beta ^2 h_{0,1}^3 \left(\left(h_{0,1}'\right){}^2-2\right)+8 h_{0,1} h_{0,1}' h_{2,3}'-4 h_{2,3} \left(\left(h_{0,1}'\right){}^2+1\right)}{2 h_{0,1}^2}\\
h_{3,4}''&=&\frac{M h_{0,1}^4 \left(4-3 \left(h_{0,1}'\right){}^2\right)-8 h_{0,1} h_{0,1}' h_{3,4}'+4 h_{3,4} \left(\left(h_{0,1}'\right){}^2+1\right)}{2 h_{0,1}^2}
\\
g_{2,3}''&=&-\frac{4 g_{2,3}' h_{0,1}'}{h_{0,1}}+\frac{2 g_{2,3} \left(\left(h_{0,1}'\right){}^2+1\right)}{h_{0,1}^2} -6 h_{0,1} h_{0,1}'
\\
g_{3,3}''&=&-\frac{4 g_{3,3}' h_{0,1}'}{h_{0,1}}+\frac{2 g_{3,3} \left(\left(h_{0,1}'\right){}^2+1\right)}{h_{0,1}^2}+\frac{2  \left(-3 \sigma  h_{0,1}^3 h_{0,1}'-2 h_{0,1}^4 \left(h_{0,1}'\right){}^2+h_{0,1}^4\right)}{h_{0,1}^2}
\end{eqnarray}
\begin{eqnarray}
\alpha(\sigma)&=&\frac{{\mathcal{E}} }{h_{0,1}}+\Omega ^2 \left(-\frac{{\mathcal{E}} ^3 h_{2,3}}{h_{0,1}^4}-\frac{{\mathcal{E}} ^2 h_{2,2}}{h_{0,1}^3}-\frac{{\mathcal{E}}  h_{2,1}}{h_{0,1}^2}\right)-\Omega^3 \frac{{\mathcal{E}} ^4 h_{3,4}}{h_{0,1}^5}
\\
\beta(\sigma)&=&-\frac{{\mathcal{E}}^3 \Omega ^2 g_{2,3} \mathbb{H}_{\text{rx}}}{h_{0,1}^4}\sin\delta-\frac{{\mathcal{E}} ^3 \Omega ^3 g_{3,3} \mathbb{H}_{\text{rx}}}{h_{0,1}^4}\cos\delta~,
\end{eqnarray}
where ${\mathcal{E}}=\frac{\epsilon}{\Omega}$ is a scaled cut-off.

\section*{B. Detailed Calculation for (\ref{Areg})}

Since $\alpha(\sigma)$ and $h_{0,1}$ are even functions and $g_{2,3}$ is an odd function,
\begin{eqnarray}
&&-\frac{\lambda\, \mathbb{H}_{rx}}{\Omega}\cos\delta\,\int_{-\sigma_{max}^{(0)}}^{\sigma_{max}^{(0)}} d\sigma \int_{\alpha(\sigma)}^R d\rho\,\frac{\rho\, \Omega^2 \left(2 g_{2,3} \left(h_{0,1}'^2+1\right)-h_{0,1} g_{2,3}' h_{0,1}'\right)}{h_{0,1}^3 \sqrt{h_{0,1}'^2+1}}=0~~.
\end{eqnarray}
By the same reason,
\begin{eqnarray}
&&\frac{\lambda\, \mathbb{H}_{rx}}{\Omega}\int_{-\sigma_{max}^{(0)}}^{\sigma_{max}^{(0)}} d\sigma \int_{\alpha(\sigma)}^R d\rho\,\rho\left(\frac{\Omega^2 h_{0,1}'\cos\delta}{\sqrt{h_{0,1}'^2+1}}+ \frac{\Omega^3 \left(\sigma  h_{0,1}'-h_{0,1}\right) \sin\delta}{\sqrt{h_{0,1}'^2+1}}\right) \nonumber\\&&~~~~=\frac{\lambda\, \mathbb{H}_{rx}}{\Omega}\sin\delta\int_{-\sigma_{max}^{(0)}}^{\sigma_{max}^{(0)}} d\sigma \int_{\alpha(\sigma)}^R d\rho\,\rho\left( \frac{\Omega^3 \left(\sigma  h_{0,1}'-h_{0,1}\right) }{\sqrt{h_{0,1}'^2+1}}\right)~~.
\end{eqnarray}
In addition,
\begin{eqnarray}
&&\frac{\lambda}{\Omega}\left( \sigma^{(1)}_{max}-\sigma^{(1)}_{min} \right)\left(\int_{\alpha(\sigma)}^{R}d\rho\left(  \sum_{n=0}^{3} \mathcal{L}^{0th}_{(n)} \Omega^n \right)_{w_0} \right)_{\sigma = \sigma^{(0)}_{max}}
\nonumber\\~~&&=\frac{\lambda}{\Omega}\left( -
\frac{\beta  \left(-\sigma ^0{}_{\max }\right)+\beta \left( \sigma ^0{}_{\max } \right) }{\alpha '\left(\sigma ^0{}_{\max }\right)} \right) \left(\int_{\alpha(\sigma)}^{R}d\rho\left(  \sum_{n=0}^{3} \mathcal{L}^{0th}_{(n)} \Omega^n \right)_{w_0} \right)_{\sigma = \sigma^{(0)}_{max}} \nonumber\\
&&=\frac{\lambda}{\Omega} \left( - \frac{2 g_{3,3} \mathcal{E}^2}{h_{0,1}' h_{0,1}^2 } \right)_{\sigma = \sigma^{(0)}_{max} } \mathbb{H}_{rx} \Omega^3  \left(\int_{\alpha(\sigma)}^{R}d\rho\left(  \sum_{n=0}^{3} \mathcal{L}^{0th}_{(n)} \Omega^n \right)_{w_0} \right)_{\sigma = \sigma^{(0)}_{max}}\nonumber\\
&&\sim \frac{\lambda}{\Omega} \left( - \frac{2 g_{3,3} \mathcal{E}^2}{h_{0,1}' h_{0,1}^2 } \right)_{\sigma = \sigma^{(0)}_{max} } \mathbb{H}_{rx} \Omega^3 \left( \frac{\sqrt{h_{0,1}'(\sigma ){}^2+1} \left(\log (R)-\log \left(\frac{\mathcal{E}}{h_{0,1}(\sigma )}\right)\right)}{h_{0,1}(\sigma ){}^2} \right)_{\sigma = \sigma^{(0)}_{max} }\nonumber\\
&&= \mathcal{O}\left( \Omega^3 \right)~~,
\end{eqnarray}
because $h_{0,1}\left(\sigma_{max}^{(0)}\right)=R/\mathcal{E}+\mathcal{O}\left(\Omega\right)$.
The last term in (\ref{Areg}) is given by
\begin{eqnarray}
&&-\frac{\lambda}{\Omega} \int_{-\sigma_{max}^{(0)}}^{\sigma_{max}^{(0)}}d\sigma \beta(\sigma) \left(  \sum_{n=0}^{3} \mathcal{L}^{0th}_{(n)} \Omega^n \right)_{\rho=\alpha(\sigma)} \nonumber\\ &&=\left(-\frac{\lambda}{\Omega}\right) \mathcal{E}^2 \int_{-\sigma_{max}^{(0)}}^{\sigma_{max}^{(0)}}d\sigma \left( -\frac{\mathbb{H}_{rx} \Omega^3 g_{3,3}(\sigma ) \sqrt{h_{0,1}'(\sigma )^2+1}}{h_{0,1}(\sigma ){}^5} \right) + \mathcal{O}\left( \Omega^3\right)~~,
\end{eqnarray}
where the integration is finite and so this term is quadratic in $\epsilon$. We may ignore this.

\providecommand{\href}[2]{#2}

\end{document}